\pdfoutput=1
\documentclass[12pt]{article}
\usepackage{multirow}
\usepackage{amsthm,amsmath,amsfonts,natbib}
\usepackage{verbatim}
\usepackage{graphicx}
\usepackage{caption}
\usepackage{tabu}
\usepackage{booktabs}
\usepackage[font=footnotesize]{caption}
\usepackage{subcaption}
\usepackage{bm}
\usepackage{bbm}
\usepackage{xcolor}
\usepackage{bbm}
\usepackage{float}
\floatstyle{plaintop}
\restylefloat{table}
\RequirePackage[colorlinks,citecolor=blue,urlcolor=blue]{hyperref}
\usepackage{natbib}
\usepackage{url} 

\newcommand{\blind}{1}

\begin{document}

\def\spacingset#1{\renewcommand{\baselinestretch}%
{#1}\small\normalsize} \spacingset{1}


\if1\blind
{
  \title{\bf A semiparametric approach for bivariate extreme exceedances}
  \author{Manuele Leonelli\\
    School of Mathematics and Statistics, University of Glasgow\\
    and \\
    Dani Gamerman \\
    Instituto de Matem\'{a}tica, Universidade Federal do Rio de Janeiro}
  \maketitle
} \fi

\if0\blind
{
  \bigskip
  \bigskip
  \bigskip
  \begin{center}
    {\LARGE\bf Title}
\end{center}
  \medskip
} \fi

\bigskip
\begin{abstract}
Inference over tails is performed by applying only the results of extreme value theory. Whilst such theory is well defined and flexible enough in the univariate case, multivariate inferential methods often require the imposition of arbitrary constraints not fully justifed by the underlying theory. In contrast, our approach uses only the constraints imposed by theory. We build on previous, theoretically justified work for marginal exceedances over a high, unknown threshold, by combining it with flexible, semiparametric copulae specifications to investigate extreme dependence. Whilst giving probabilistic judgements about the extreme regime of all marginal variables, our approach formally uses the full dataset and allows for a variety of patterns of dependence, be them extremal or not. A new probabilistic criterion quantifying the possibility that the data exhibits asymptotic independence is introduced and its robustness empirically studied. Estimation of functions of interest in extreme value analyses is performed via MCMC algorithms. Attention is also devoted to the prediction of new extreme observations. Our approach is evaluated through a series of simulations, applied to real data sets and assessed against competing approaches. Evidence demonstrates that the bulk of the data does not bias and improves the inferential process for the extremal dependence.
\end{abstract}

\noindent%

\spacingset{1.45} 
\section{Introduction}
\label{sec:intro}
Precise knowledge of the tail behaviour of a distribution as well as predicting capabilities about the occurrence of extremes are fundamental in many areas of applications, as for instance environmental sciences and finance amongst many. Evidence points out to an increasing trend of such extreme events in environmental applications with associated economic and insurance losses growing dramatically \citep{Salvatori2007}. In most cases the analysis of such extreme events is inherently multivariate. Interest is then on the concomitant observation of extremes on a number of variables. For instance, the effects on the human respiratory system are particularly dramatic for exposition to high concentrations of both ozone O$_3$ and nitrogen dioxide NO$_2$. 

Since  standard statistical methods do not guarantee precise extrapolation towards the tail of the distribution, a variety of methods tailored to inference about tails have been introduced under the general name of \textit{extreme value theory}.  Whilst univariate models can be faithfully applied in most applications, since their underlying assumptions are  flexible enough to be met in practice, the application of multivariate methods often requires the imposition of ad-hoc assumptions about the asymptotic dependence structure: for instance by excluding the possibility of asymptotic independence. Furthermore, the application of such methods requires the arbitrary selection of datapoints considered \lq\lq{e}xtreme\rq\rq{}, usually selected as those that exceed a fixed  threshold. However, this choice can greatly affect the inferential process \citep{Scarrott2012}. To overcome these difficulties a new easily interpretable, flexible approach is proposed here to investigate both marginal and joint extreme behaviours that formally uses in a model-based fashion the full dataset. This combines some fairly new methodology for univariate extremes justified by the the asymptotic theory for tails, with a flexible semiparametric dependence structure definition which does not require any assumption about the asymptotic dependence decay. Empirical evidence demonstrates in Section \ref{sec:uff} below that in the applications considered the bulk of the data does not bias our inferential ascertainment of the asymptotic dependence structure.

Inference is carried out within the \textit{Bayesian} paradigm using the \textit{MCMC} machinery \citep{Gamerman2006}, enabling us to straightforwardly deliver a wide variety of estimates and predictions of quantities of interest, e.g. high quantiles. Although our methods could be straightforwardly extended to a more general nonparametric approach, we are able to demonstrate below that our simpler and computationally less intensive methodology can capture diverse patterns of dependence, be them extremal or not. 

In this work, as often in the literature, we focus on problems where extreme behaviour is of interest on the right tail  only. However our approach could be easily extended to handle situations where interest is on both tails \citep[e.g.][]{Scarrott2012}.

Before formally defining our approach, 
both univariate and multivariate extreme value theory and copulae functions are briefly introduced to highlight the relevance and the novelty of our methodology.

\subsection{Univariate extreme value theory}
A common approach to model extremes, often referred to as \textit{peaks over threshold} (POT),  studies the exceedances over a threshold. A key result to apply this methodology is due to \citet{Pickands1975} which states that if a random variable X with endpoint $x_e$ is in the  domain of attraction of a generalized extreme value distribution \citep[see e.g.][]{Beirlant2004} then $\lim_{u\rightarrow x_e}\mathbb{P}(X\leq x+u|X>u)=P(x)$, where $P$ is the distribution function (df) of the  \textit{generalized Pareto distribution} (GPD). The df $P$ is defined as
\begin{equation*}
P(x|\xi,\sigma,u)=\left\{
\begin{array}{ll}
1-\left(1+\xi\frac{x-u}{\sigma}\right)^{-1/\xi},& \mbox{if } \xi\neq 0,\\
1-\exp\left(-\frac{x-u}{\sigma}\right), & \mbox{if }\xi=0,
\end{array}
\right.
\end{equation*}
for  $u,\xi\in\mathbb{R}$ and $\sigma\in\mathbb{R}_{+}$, where the support is $x{\geq u}$ if $\xi\geq 0$ and $0\leq x\leq u-\sigma/\xi$ if $\xi<0$. Therefore, the GPD is bounded if $\xi<0$ and unbounded from above if $\xi\geq 0$. The application of this result in practice entails first the selection of a threshold $u$ beyond which the GPD approximation appears to be tenable and then the fit of a GPD over data points that exceed the chosen threshold.

The POT approach has two serious drawbacks. First, only a small subset of the data points, those beyond the chosen threshold, are formally retained in a model-based approach during the inferential process. Thus parameter estimates may not be reliable when the number of data points is small. Second, the choice of the threshold over which to fit a GPD is arbitrary. Although  tools to guide this choice exist \citep[e.g.][]{Davison1990}, inference can greatly vary for different thresholds \citep{Einmahl2009a,Scarrott2012}.

To overcome these deficiencies, a variety of models called \textit{extreme value mixture models} \citep[][]{Scarrott2012} have been recently defined to formally take into account the full dataset and not require a fixed threshold. These  combine a flexible model for the bulk of the data points, those below the threshold, a formally justifiable model for the tail and uncertainty measures for the threshold. A building block of our approach is the  \textit{MGPD} extreme value mixture model of \citet{Nascimento2012}.

\vspace{0.15cm}

\noindent\textbf{The MGPD model}

\vspace{0.15cm}

\noindent The flexible MGPD model consists of a finite mixture of gamma distributions for the bulk coupled with a GPD for the tail.   The parametrization of the gamma suggested in \citet{Wiper2001} in terms of shape, $\eta$, and mean, $\mu$, parameters is used to avoid identifiability issues  \citep[e.g.][]{Richardson1997}. Its density, $g$, is $g(x|\mu,\eta)=\Gamma(\eta)^{-1}\left(\eta/\mu\right)^{\eta}x^{\eta-1}\exp\left(-\eta x/\mu\right)$ and its df is denoted by $G$.

\begin{figure}
\begin{center}
\includegraphics[scale=0.5]{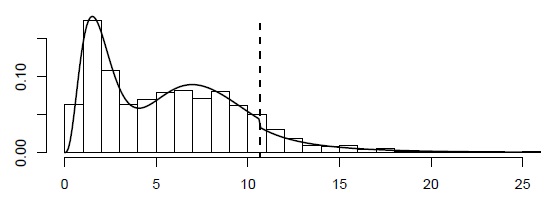}
\end{center}
\vspace{-1cm}
\caption{\footnotesize{Example of a MGPD density fit consisting of a mixture of 2 gammas for the bulk. }\label{fig:ex}}
\end{figure}

 A finite mixture of these distributions is defined next. For $n\in\mathbb{N}$, let $[n]=\{1,\dots,n\}$. The density $h$ and the df $H$ of a finite mixture of $n$ gammas are formally defined as
\begin{equation}
\label{eq:mixture}
h(x|\bm{\mu},\bm{\eta},\bm{w})=\sum_{i\in[n]}w_ig(x|\mu_i,\eta_i), \hspace{1cm} H(x|\bm{\mu},\bm{\eta},\bm{w})=\sum_{i\in[n]}w_iG(x|\mu_i,\eta_i),
\end{equation}
where $\bm{\mu}=(\mu_i)_{i\in[n]}$, $\bm{\eta}=(\eta_i)_{i\in[n]}$, $\bm{w}=(w_i)_{i\in[n]}$ and $\bm{w}$ is such that $w_i\geq 0$ and $\sum_{i\in[n]}w_i=1$.

The density $f$ of an MGPD then consists of a mixture of gamma densities $h$ for the bulk and a GPD density $p$ for the right tail. Formally, 
\begin{equation*}
f(x|\Theta)=\left\{
\begin{array}{ll}
h(x|\bm{\mu},\bm{\eta},\bm{w}), & \mbox{if } x\leq u,\\
\left(1-H(u|\bm{\mu},\bm{\eta},\bm{w})\right)p(x|\xi,\sigma,u), & \mbox{if } x> u,
\end{array}
\right.
\end{equation*}
where $\Theta=\{\bm{\mu},\bm{\eta},\bm{w},\xi,\sigma,u\}$. An example of an MGPD density fitting simulated data is presented in Figure \ref{fig:ex}, where it is clearly discernible that the bulk of the distribution consists of a mixture of 2 gammas, whilst beyond the threshold  the density has GPD decay.

The df of a MGPD $F$ is similarly defined in a piece-wise fashion. Whilst below the threshold $u$ this is the df of the mixture of gammas $H$, over the threshold, i.e. for $x>u$, it can be written as $F(x|\Theta)=H(u|\bm{\mu},\bm{\eta},\bm{w})+\left(1-H(u|\bm{\mu},\bm{\eta},\bm{w})\right)P(x|\xi,\sigma,u)$.

A great advantage of the MGPD model is that high quantiles beyond the threshold, i.e. $q$ values such that $\mathbb{P}(X>q|\Theta)=1-p$ for $p$ close to 1, have a closed-form expression. Specifically, this is a function $q$ of both the probability $p$ and the parameter $\Theta$ defined as 
\begin{equation*}
q(p|\Theta)=u+\frac{\sigma}{\xi}\left(\left(1-\frac{p-H(u|\bm{\mu},\bm{\eta},\bm{w})}{1-H(u|\bm{\mu},\bm{\eta},\bm{w})}\right)^{-\xi}-1\right).
\end{equation*} 

\citet{Nascimento2012} demonstrated that the MGPD can outperform standard POT models in situations where determination of the threshold is difficult. So nothing is lost using this approach instead of considering only the extreme points as in the standard POT method. The MGPD also provides better estimates than a standard nonparametric mixture model with an arbitrary large number of gamma components. Furthermore, a finite mixture is sufficient to model the bulk of the distribution since the weights of the required gamma components only are non-zero \citep{Nascimento2012}.  

\subsection{Multivariate extreme value theory}
\label{sec:multi}
Modelling approaches for multivariate extremes rely on limiting results of componentwise maxima and are mainly due to \citet{deHaan1977}. One of these limiting results is briefly discussed next and refer to \citep[see e.g.] [for a comprehensive review]{Beirlant2004}.

Let $\bm{X}_1,\dots,\bm{X}_{n}\in\mathbb{R}^{d}_+$, where $\bm{X}_i=(X_{ij})_{j\in[d]}$, be independent and identically distributed random vectors with marginal unit Fr\'{e}chet distributions with dfs $\exp(-1/x)$, $x\in\mathbb{R}_{+}$. If the componentwise maximum $\bm{M}_n=\left(\max_{i\in[n]}X_{ij}\right)_{j\in[d]}$ converges in distribution as $n\rightarrow \infty $ to a non-degenerate df $E$, then 
$E(\bm{x})=\exp(-V(\bm{x}))$,  where  $V(\bm{x})=d\int_{\mathcal{S}_d}\max_{i\in[d]}\omega_i/x_iH(\textnormal{d}\bm{\omega})$, $\bm{w}=(w_i)_{i\in[d]}$, $\mathcal{S}_d$ is the d-dimensional unit simplex, i.e. $\mathcal{S}_d=\{\bm{\omega}:\omega_i\geq 0, \sum_{i\in[d]}\omega_i=1\}$, and $H$ is a probability measure on $\mathcal{S}_d$ satisfying the \lq\lq{m}ean\rq\rq{} constraint $\int_{\mathcal{S}_d}\omega_iH(\textnormal{d}\bm{\omega})=d^{-1}$. The function $V$ is  called \textit{exponent measure}, whilst $H$ is the \textit{spectral measure}. The df $E$ is called \textit{multivariate extreme value distribution} (MEVD).

The main point here is that the limiting distribution of componentwise  maxima cannot be described in a parametric closed form, but consists of a nonparametric family characterized by the spectral functions respecting the \lq\lq{m}ean\rq\rq{} constraint. The generality of this result has lead to the definition of a variety of approaches to model multivariate extreme observations. We can broadly identify three different strategies:
\begin{itemize}
\item define a parametric submodel for either the exponent measure \citep{Coles1991,Coles1994,Jaruskova2009} or the spectral measure \citep{Ballani2011,Boldi2007,Cooley2010};
\item model in a nonparametric fashion the class of MEVD distributions \citep{Einmahl2009,Guillotte2011};
\item construct models based on alternative theoretical justifications \citep{Bortot2000,Carvalho2014,Ramos2009,Heffernan2016}.
\end{itemize}

In all cases, data is usually transformed via the empirical df into Fr\'{e}chet margins and then some of the data points, those considered \lq\lq{e}xtreme\rq\rq{}, are formally retained for inference. Having already discussed the difficulty of assessing such a threshold in the univariate case, the identification of extreme data points becomes even more critical in multivariate applications since there is no unique definition of threshold.

\begin{figure}
\begin{center}
\begin{subfigure}{0.23\textwidth}
\centering
\includegraphics[scale=0.35]{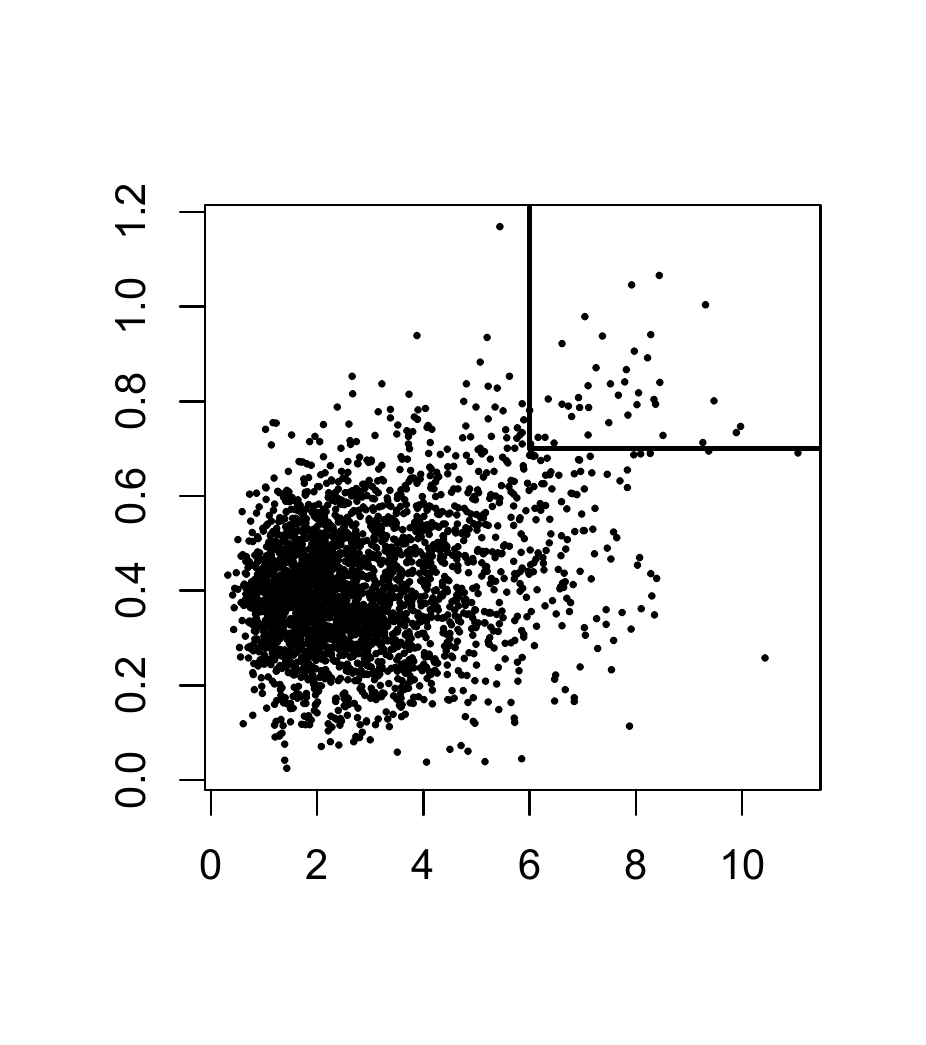}
\vspace{-0.5cm}
\caption{\footnotesize{\label{fig:upper}}}
\end{subfigure}
\begin{subfigure}{0.23\textwidth}
\centering
\includegraphics[scale=0.35]{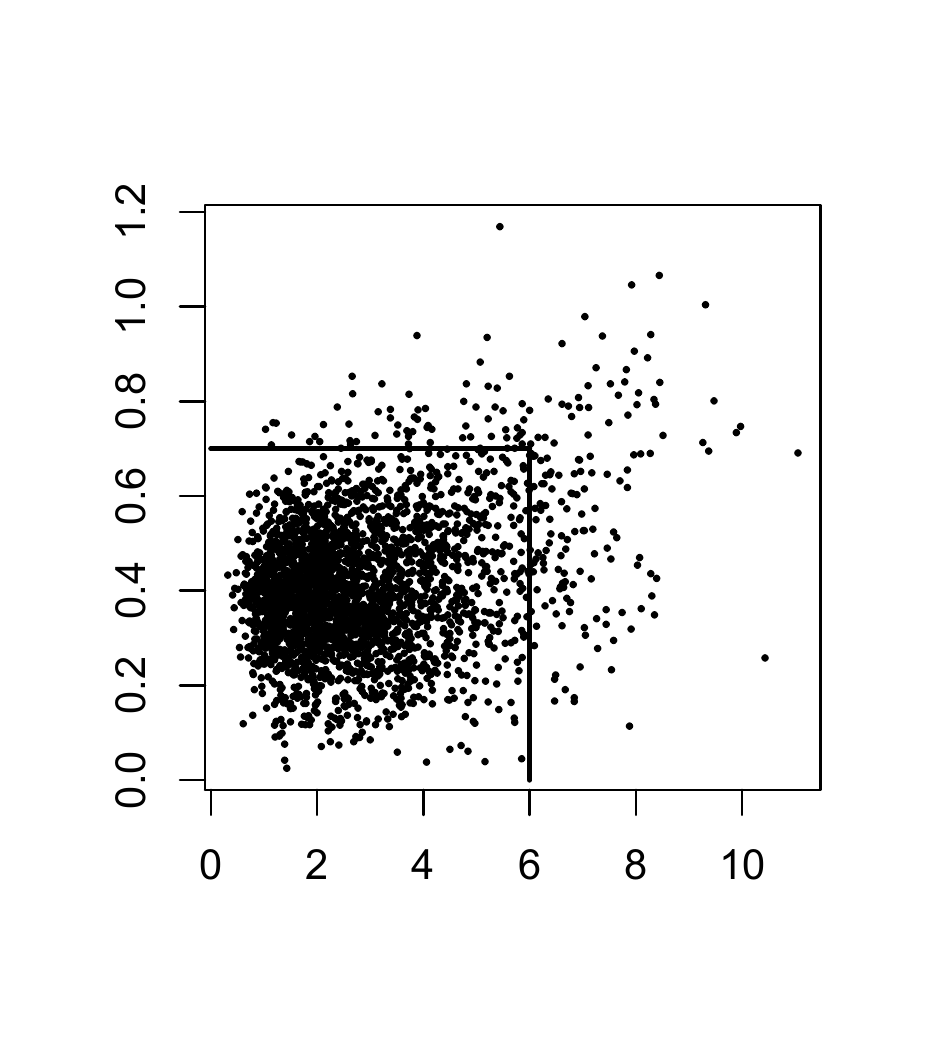}
\vspace{-0.5cm}
\caption{\footnotesize{\label{fig:lower}}}
\end{subfigure}
\begin{subfigure}{0.23\textwidth}
\centering
\includegraphics[scale=0.35]{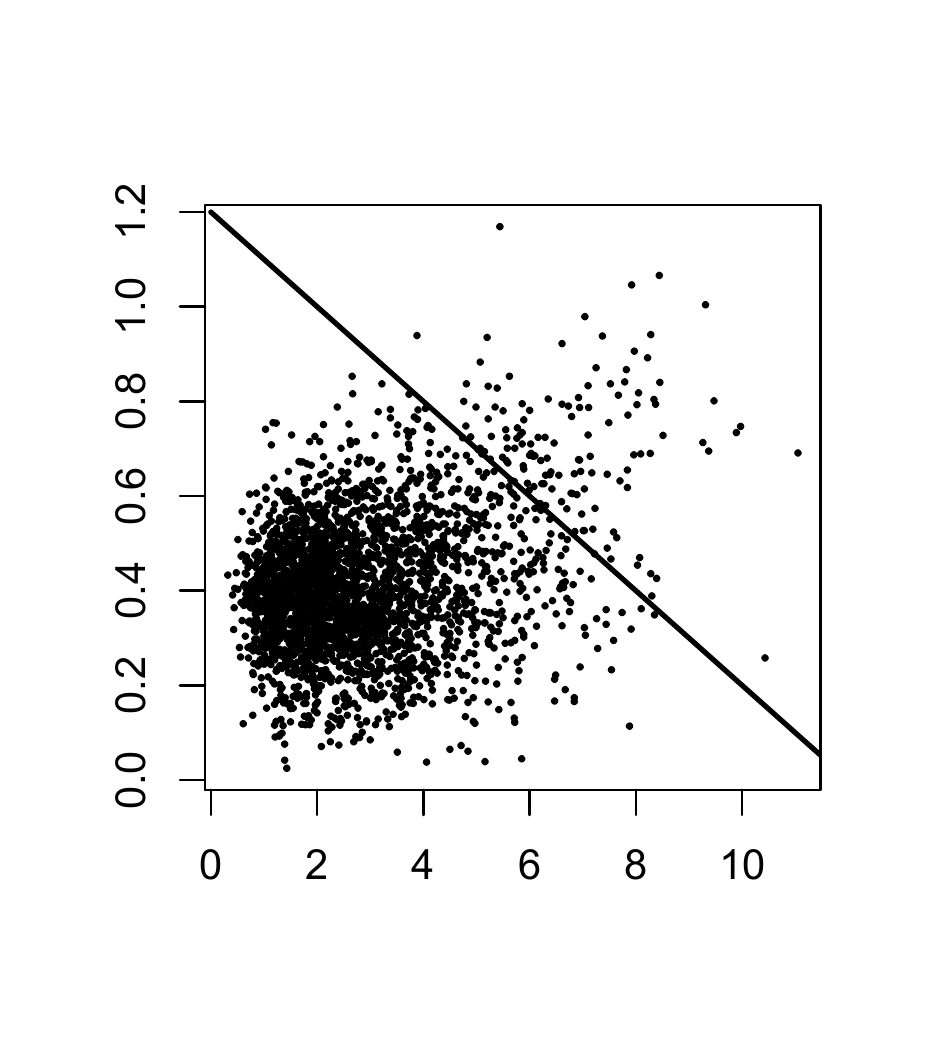}
\vspace{-0.5cm}
\caption{\footnotesize{\label{fig:bend}}}
\end{subfigure}
\begin{subfigure}{0.23\textwidth}
\centering
\includegraphics[scale=0.35]{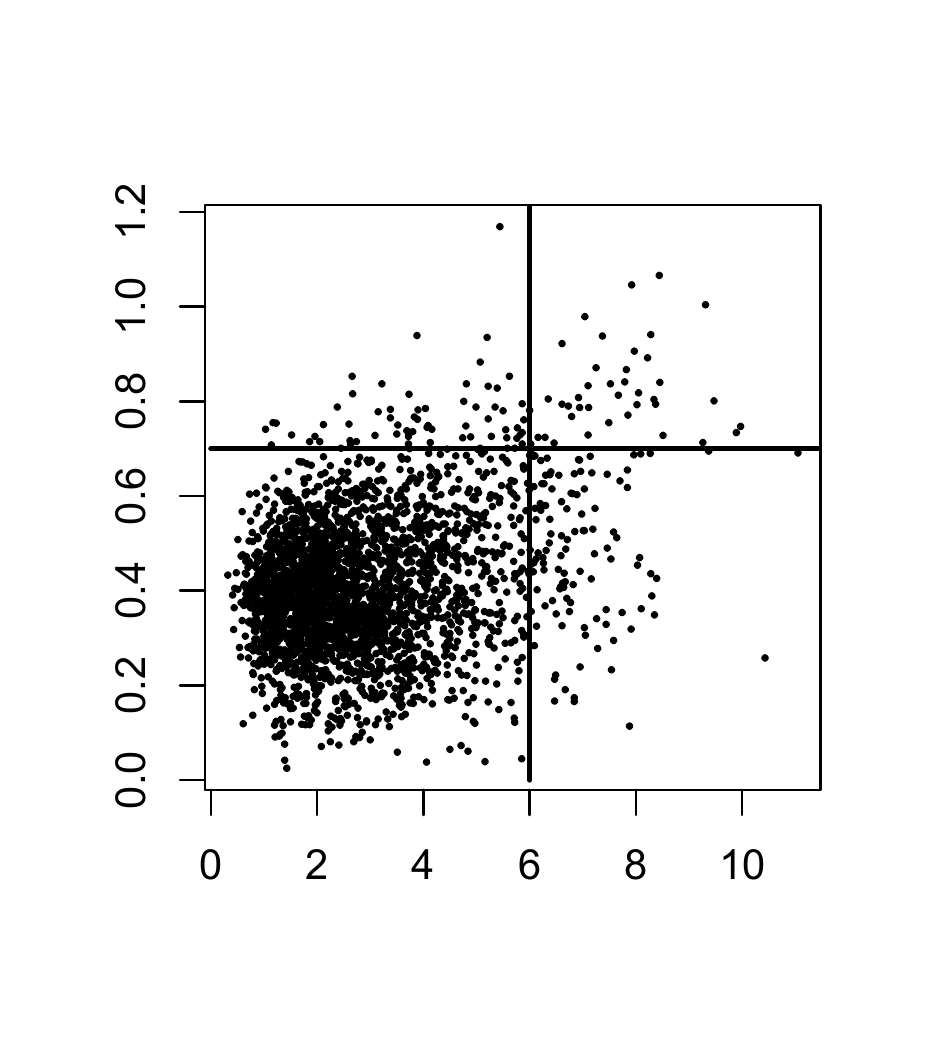}
\vspace{-0.5cm}
\caption{\footnotesize{\label{fig:censored}}}
\end{subfigure}
\end{center}
\vspace{-0.75cm}
\caption{\footnotesize{Examples of bivariate threshold choices.}\label{fig:thresholds}}
\end{figure}

To illustrate this, consider the different bivariate threshold choices in Figure \ref{fig:thresholds}. Figures \ref{fig:upper} and \ref{fig:lower} state that an observation is extreme if it is  beyond the threshold in all or in at least one component, respectively. These thresholds are usually utilized when estimating contemporaneously marginal and joint features of the data. The threshold in Figure \ref{fig:bend} describes as extreme an observation such that the sum of its components is larger than a specified value and is often used when only modelling dependence. The last threshold in Figure \ref{fig:censored} is associated to the so called censored approach: an observation below a marginal threshold in any component is supposed to be censored at the threshold. 

Although the theoretical limiting result of maxima can be expected to hold in the region specified by the threshold in Figure \ref{fig:upper}, all other thresholds are more commonly utilized to increase the sample size effectively retained for inference. Furthermore, the choice of such thresholds is often driven by the type of analysis required or computational simplifications. A flexible method that takes into account the full dataset is developed here to avoid making the arbitrary choices of thresholds location and type.

Furthermore, MGEV distributions assume a constant degree of dependence between pairs of rvs.  However, in many practical applications dependent variables are observed to be asymptotically independent and many commonly used distributions exhibit this behavior: e.g. the bivariate normal with correlation $\rho\in(-1,1)$, $\rho\neq 0$. Due to a result of \citet{Berman1961}, multivariate extreme independence can be assessed by investigating all pairs of random variables. We thus focus on bivariate vectors. \citet{Sibuya1960} proved that two random variables $X_1$ and $X_2$ with dfs $F_1$ and $F_2$  are asymptotically independent iff the \textit{coefficient of asymptotic independence} $\chi$ is equal to zero, where $\chi=\lim_{u\rightarrow 1}\chi(u)$, and $\chi(u)=\mathbb{P}(F_1(X_1)>u|F_2(X_2)>u)$. For instance, for a bivariate MGEV distribution $\chi=0$ iff $X_1$ and $X_2$ are independent, whilst $\chi=0$ for any bivariate Gaussian with dependence $\rho\neq |1|$. To address this deficiency of the MGEV distribution, novel extreme models that can take into account asymptotic dependence and independence have been proposed \citep{Heffernan2004,Ramos2009,Heffernan2016}.

Since $\chi=0$ for all asymptotically independent bivariate vectors, this criterion does not provide information about the relative strength of dependence for independent extremes.  \citet{Coles1999} defined the coefficient of subasymptotic dependence $
\bar{\chi}=\lim_{u\rightarrow 1}\bar{\chi}(u)$, where 
\[
\bar{\chi}(u)=
\frac{2\log(\mathbb{P}(F_1(X_1)>u))}{\log(\mathbb{P}(F_1(X_1)>u,F_2(X_2)>u))}-1.
\]
If $\bar{\chi}=1$ then $X_1$ and $X_2$ are asymptotically dependent, whilst if $\bar{\chi}\in(-1,1)$ then $X_1$ and $X_2$ are asymptotically independent. The strength of dependence increases with $\bar{\chi}$.

\subsection{Copulae}
\label{sec:copula}
Having chosen to model the marginals as MGPDs, a tool to construct multivariate distributions with such given margins is needed. Copulae are flexible functions to model complex relationships in a simple way. These only model the dependence structure of a random vector and allow for marginals to be defined separately \citep[see][for a review]{Nelsen2006}.

For a random vector $\bm{X}=(X_i)_{i\in[d]}$ with df $F$, whose margins have dfs $F_i$, $i\in[d]$,  a \textit{copula} $C$ is defined as a function $C:[0,1]^d\rightarrow [0,1]$ such that
$
F(\bm{x})=C(F_1(x_1),\dots, F_d(x_d))$. 
\citet{Sklar1959}  proved that such a $C$ linking marginal and joint distributions always exists. Notice that $C$ is a df itself and as such possesses a density $c$ called \textit{copula density} and defined as $c(\bm{v})=\partial C(\bm{v})/\partial \bm{v}$, for $\bm{v}\in[0,1]^d$. Thus the density of $\bm{X}$ equals $
f(\bm{x})=c(F_1(x_1),\dots, F_d(x_d))\prod_{i\in[d]}f_i(x_i)$,
 where $f_i$  and $f$ are the densities of $X_i$ and $\bm{X}$ respectively. 
 
Copulae and finite mixture models have recently been combined \citep[e.g.][]{Kim2013} to depict an even wider variety of patterns of dependence. Formally, a mixture of $n$ copulae $C_i$, $i\in[n]$, is  defined as $\sum_{i\in[n]}w_iC_i(F_1(x_1),\dots,F_d(x_d))$, where $\sum_{i\in[n]}w_i=1$ and $w_i\geq 0$.

\subsection{Outline of the paper}
Our approach and inferential routines are next described  in Section \ref{sec:models}. Section \ref{sec:simulations} presents a simulation study to both investigate their performance and address the issue of model choice.  In Section \ref{sec:applications}  our methodology is applied to two real-world applications: river flows in Puerto Rico and NO$_3$/O$_2$ concentrations in the city of Leeds. We conclude with a discussion.

\section{The semiparametric approach}
\label{sec:models}

\subsection{Likelihood}

For each marginal, an MGPD with density and df $f_i$ and $F_i$ respectively and parameters $\Theta_i=\{\bm{w}_i,\bm{\eta}_i,\bm{\mu}_i,\xi_i,\sigma_i,u_i\}$ is used, where $\bm{w}_i=(w_{ij})_{j\in[n_i]}$, $\bm{\eta}_{i}=(\eta_{ij})_{j\in[n_i]}$ and $\bm{\mu}_{i}=(\mu_{ij})_{j\in[n_i]}$ are the parameters of a mixture of $n_i$ gammas as in equation (\ref{eq:mixture}). The dependence structure is modelled by a mixture of $n$ copulae $C_i$ with weights  $\bm{w}=(w_i)_{i\in[n]}$ and parameter set $\Theta_{D_i}$, $i\in[n]$. Letting $\Theta=\{\bm{w},\Theta_{D_i},\Theta_j:i\in[n],j\in[d]\}$, our df $F$  is given by
\begin{equation*}
\label{eq:df} 
 F(\bm{x}|\Theta)=\sum_{i\in[n]}w_iC_i(F_1(x_1|\Theta_1),\dots,F_d(x_d|\Theta_d)|\Theta_{D_i})
\end{equation*} 
  and its \textit{density} $f$ equals
\begin{equation}
\label{eq:model}
f(\bm{x}|\Theta)=\sum_{i\in[n]}w_ic_i(F_1(x_1|\Theta_1),\dots,F_d(x_d|\Theta_d)|\Theta_{D_i})\prod_{i\in[d]}f_i(x_i|\Theta_i),
\end{equation}
where $c_i$ is the associated copula density, $i\in[d]$.

Although our approach does not require any restriction on the chosen copulae, in this work  mixtures of elliptical copulae are used: more specifically, Gaussian \citep{Song2000}, T \citep{Demarta2005},  skew-normal \citep{Wu2014} and  skew-T \citep{Smith2012} copulae. Furthermore  all mixture components are assumed to belong to the same family, e.g. Gaussian. Such mixtures have the very convenient property of a known asymptotic behavior: whilst mixtures of Gaussians and skew-normals have  asymptotically independent extremes,  Ts and skew-Ts exhibit extreme dependence \citep{Bortot}.

Consider now bivariate vectors only. The specific form of our densities follows by substituting $c_i$ in equation (\ref{eq:model}) with the expressions in the Supplementary Material. Simulation studies showed that, for full parameter identification, restrictions need to be imposed on the likelihood in equation (\ref{eq:model}). Whilst for mixtures of Gaussian copulae no constraints are imposed, for the other  mixtures  the following is assumed:
\begin{itemize}
\item for T-copulae all components have the same number of degrees of freedom in $\mathbb{R}_+$;
\item for skew-Normal copulae all components have the same skewness parameters;
\item for skew-T copulae one single component with integer degrees of freedom.\footnote{This greatly speeds up computations using the formulae of \citet{Dunnett1954}}
\end{itemize} 

As well as having closed form expressions for marginal quantiles, bivariate quantiles can be easily deduced in our models. However, these are not uniquely defined since there are infinitely many pairs  $(x_1,x_2)$ such that  $\mathbb{P}(X_1>x_1,X_2>x_2|\Theta)$ is equal to a specified number. Thus we look at pairs $(x_1,x_2)$ and compute the associated probability of \textit{joint exceedance} $\mathbb{P}(X_1>x_1,X_2>x_2|\Theta)$. This is a function $E$ of $(x_1,x_2)$ and  $\Theta$ defined as
\begin{equation}
\label{eq:joint}
E(x_1,x_2|\Theta)=1-F_1(x_1|\Theta_1)-F_2(x_2|\Theta_2)+\sum_{i\in[n]}w_iC_i(F_1(x_1|\Theta_1),F_2(x_2|\Theta_2)|\Theta_{D_i}).
\end{equation}

Similarly, our approach leads to closed-form expressions for the probabilities $\chi(u|\Theta)$ and $\bar{\chi}(u|\Theta)$ appearing in the coefficients of asymptotic and subasymptotic independence respectively. This is because, for instance, $\chi(u|\Theta)=\mathbb{P}(F_1(X_1|\Theta_1)>u)/E(F_1^{-1}(u|\Theta_1),F_2^{-1}(u|\Theta_2),\Theta)$ and these two probabilities have closed form expressions.
   
\subsection{Prior distribution}
\label{sec:priors}
Our approach is completed by the introduction of a prior distribution, defined by considering separate blocks of parameters.

For each of the marginal components $\Theta_i$  the priors specified in \citet{Nascimento2012} are used. Specifically, for the $i$-th marginal component, to each $\eta_{ij}$ a gamma prior with shape $c_{ij}$ and mean $d_{ij}$ is assigned, where these parameters may be chosen to achieve a large prior variance.  The parameter space of $\bm{\mu}_i$ is restricted to $C(\bm{\mu}_i)=\{\bm{\mu}_i: 0<\mu_{i1}<\cdots <\mu_{in_i}\}$ to address the identifiability issues of mixtures. To each $\mu_{ij}$  an inverse gamma prior with shape $a_{ij}$ and mean $b_{ij}$ is assigned, where again these parameters may be chosen to achieve a large variance. Therefore the prior for $\bm{\mu}_i$ is
$
\pi(\bm{\mu}_i)=K\prod_{j\in[n_i]}f_{IG}(\mu_{ij}|a_{ij},b_{ij})\mathbbm{1}_{C(\bm{\mu}_i)}(\bm{\mu}_i),$
where $\mathbbm{1}_{A}(x)=1$ if $x\in A$ and zero otherwise,  $f_{IG}$ is the inverse gamma density and 
$K^{-1}=\int_{C(\bm{\mu}_i)}\prod_{i\in[n_i]}f_{IG}(\mu_{ij}|a_{ij},b_{ij})\partial\bm{\mu_i}.$ The weights of the gamma mixture, $w_{ij}$, are assigned a Dirichlet $D(\bm{1}_{n_i})$ prior, where $\bm{1}_{n_i}$ is a vector of dimension $n_i$ with ones in all entries.

The prior of the threshold $u_i$ is normal as in \citet{Nascimento2012} and \citet{Behrens2004}. Care must be exercised when specifying the hyperparameters of this distribution. The mean is chosen around a high order sample statistics. The variance is chosen so that the bulk, say 95$\%$, of the prior distribution ranges roughly over data points larger than the median. These variances need to be slightly smaller than in the univariate MGPD model to ensure convergence. 

The hyperparameters above can be changed to effectively include expert prior information without affecting our inferential routines. 
 
For the shape and scale of the GPD distributions the uninformative prior of \citet{Castellanos2007} is used, defined as
$\pi(\xi_i,\sigma_i)=\sigma_i^{-1}(1+\xi_i)^{-1}(1+2\xi_i)^{-1/2}$, $i\in[d]$.

 For correlation coefficients $\rho_i$  a continuous uniform $\mathcal{U}[-1,1]$ is selected. The joint $\pi(\bm{\rho})$ is defined over a restricted space as for the mean parameters of the gamma mixtures to ensure identifiability. For skew copulae a continuous uniform $\mathcal{U}[-1+\epsilon,1-\epsilon]$ is assigned to the skewness parameters $\delta_j$, for an $\epsilon$ close to zero. The copulae mixture weights $w_i$ are given a Dirichlet $D(\bm{1}_{n})$. These priors are chosen to give uninformative prior beliefs.

For the degrees of freedom $v$ of the T-copula the uninformative prior of \citet{Fonseca2008}  is used, defined as
\begin{equation*}
\pi(v)=\left(\frac{v}{v+3}\right)^{1/2}\left(\phi\left(\frac{v}{2}\right)-\phi\left(\frac{v+1}{2}\right)-\frac{2(v+3)}{v(v+1)^2}\right)^{1/2}, \,\,\,\, v\in\mathbb{R}_+
\end{equation*}
where $\phi$ is the trigamma function. For the skew-T copula with integer degrees of freedom a zero-truncated Poisson distribution with mean 25 is used. Sensitivity studies showed that this value enabled for the identification of both low and high degrees of freedom.

The overall \textit{prior distribution} is then defined as
\begin{equation*}
\label{eq:prior}
\pi(\Theta)=\pi(\bm{w})\pi(\Theta_D)\prod_{i\in[2]}\pi(\xi_i,\sigma_i)\pi(u_i)\pi(\bm{w}_i)\pi(\bm{\mu}_i)\prod_{j\in[n_i]}\pi(\eta_{ij}),
\end{equation*}
where $\Theta_D\subseteq\{\bm{\rho},v,\delta_1,\delta_2\}$ and $
\log(\pi(\Theta_D))=\log(\pi(\bm{\rho}))+\mathbbm{1}_{\Theta_D}(v)\log(\pi(v))+\mathbbm{1}_{\Theta_D}(\delta_1)\log(\pi(\delta_1)\pi(\delta_2))$. The set $\Theta_D$ is so defined to encompass all elliptical copulae considered in this paper.

\subsection{Posterior and predictive inference}
For a sample $\bm{x}=(\bm{x}_i)_{i\in[m]}$, where $\bm{x}_i=(x_{1i},x_{2i})$, the \textit{posterior} log-density is then
\begin{equation}
\label{eq:posterior}
\log\pi(\Theta|\bm{x})=\sum_{j\in[m]}\log\Big(\sum_{i\in[n]}w_ic_i(F_1(x_{1j}|\Theta_1),F_2(x_{2j}|\Theta_2)|\Theta_{D_i})\Big)+\sum_{i\in[2]}\log\left(f_i(x_{ij}|\Theta_i)\right)+\log(\pi(\Theta)).
\end{equation}
Inference cannot be performed analytically and approximating MCMC algorithms are used. Parameters are divided into blocks and updating of the blocks follows Metropolis-Hastings steps since full conditionals have no recognizable form. Proposal variances are tuned via an adaptive algorithm as suggested in \citet{Roberts2009}.  Details  are given in the Supplementary Material. All algorithms are implemented in OX \citep{Doornik1996}. 

Most quantities of interest in the analysis of extremes, e.g. $\chi(u|\Theta)$, are highly non-linear functions of the models' parameters. Thus their posterior distribution cannot be derived analytically. However, the MCMC machinery  enables us to derive an approximated distribution for \textit{any} function of the models' parameters. For instance, for $I$ draws $\Theta^{(i)}$, $i\in[I]$, from the posterior $\pi(\Theta|\bm{x})$, the values $\chi(u|\Theta^{(i)})$ approximate the posterior distribution of $\chi(u|\Theta)$, given a sample $\bm{x}$. An estimate of the posterior mean is then $\frac{1}{I}\sum_{i\in I}\chi(u|\Theta^{(i)})$. 
  
Estimation is an important task in extreme value theory as much as the prediction of a new observation $\bm{x}_{m+1}$ given a sample $\bm{x}$. The likelihood of a new observation can be summarized by the \textit{predictive} distribution of joint exceedance $E(\bm{x}_{m+1}|\bm{x})$ given by 
\begin{equation*}
E(\bm{x}_{m+1}|\bm{x})=\int E(\bm{x}_{m+1},\Theta|\bm{x})\textnormal{d}\Theta=\int E(\bm{x}_{m+1}|\Theta)\pi(\Theta|\bm{x})\textnormal{d}\Theta=\mathbb{E}_{\Theta|\bm{x}}(E(\bm{x}_{m+1}|\Theta)).
\end{equation*}
This corresponds to the expectation of equation (\ref{eq:joint}) with respect to the posterior $\pi(\Theta|\bm{x})$. This expectation cannot be computed analytically, but our Bayesian approach enables us to derive an approximated Monte Carlo estimate equal to $\frac{1}{I}\sum_{i\in[I]}E(\bm{x}_{m+1}|\Theta^{(i)})$. 

\subsection{Ascertainment of extreme independence}
\label{sec:asce}
A critical task in the analysis of extremes is the determination of the asymptotic dependence structure. However very few models are able to take into account both extreme dependence and independence, and consequently discriminate one from the other. More importantly, to our knowledge none of these can deliver a probabilistic judgement about the data exhibiting either behavior. In contrast, our semiparametric Bayesian approach enables us to introduce a new probabilistic criterion for the ascertainment of asymptotic independence based on the posterior distribution of the degrees of freedom of the T copula.
 
 \begin{figure}
\begin{center}
\includegraphics[scale=0.4]{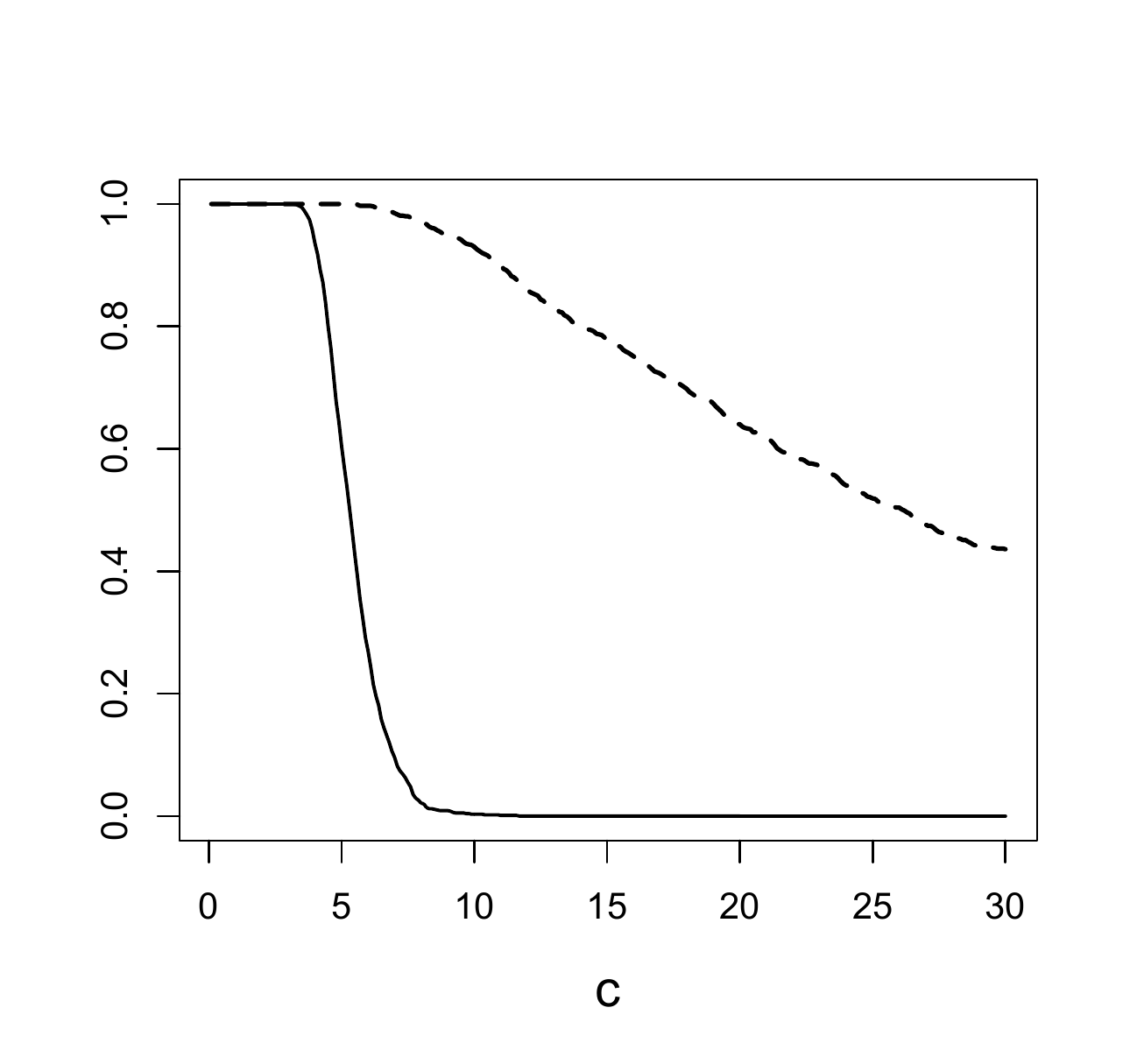}
\end{center}
\vspace{-0.9cm}
\caption{\footnotesize{$\phi(c):$ River data (full line) - Leeds data (dashed line).} \label{fig:degrees}}
\end{figure}

 Recall that for $v\rightarrow \infty$, T and skew-T copulae tend to Gaussian and skew-normal ones, respectively, and consequently large posterior estimates of the degrees of freedom may indicate asymptotically independent extremes. Thus, for a fixed $c\in\mathbb{R}_+$, we define the criterion $\phi(c)=\mathbb{P}(v\in(c,\infty)|\bm{x})$ which gives an uncertainty measure about the possibility that $\chi=0$ and thus that extremes are independent. Values of $\phi(c)$ close to zero give a strong indication of asymptotic dependence, whilst for $\phi(c)$ close to one the evidence is towards asymptotic independence. In our experience, the estimation of the number of degrees of freedom is more robust for T-copulae, possibly because not affected by prior parameters. For these mixtures a value $c=10$ seems to provide a sound uncertainty statement, as shown in Figure \ref{fig:degrees}, where the function $\phi(c)$ from the analyses carried out in  Section \ref{sec:applications} below is reported. So for instance the solid line denotes $\phi(c)$ for a dataset that exhibits dependent extremes and $\phi(c)\approx 0$ for $c\geq 10$. Thus  hereafter $\phi=\phi(10)$ denotes our summary of evidence towards asymptotic independence.

\section{Simulations}
\label{sec:simulations}
A simulation study, performed to  validate selection criteria for our mixtures, is summarized next. Importantly, this exercise enabled us to identify a variety of factors that together can provide a reliable toolkit to identify the strength of extreme dependence. 

The study consisted of 8 samples of size 1000 from a variety of dependence structures and marginals. Specifically, data was  simulated from: a mixture of 2 Gaussian copulae with MGPD margins (2G); a skew-Normal copula with MGPD margins (SN); a Morgenstern copula with lognormal-GPD margins (MO); a bilogistic copula with lognormal margins (BL); a mixture of 2 T copulae with MGPD margins (2T); a  skew-T copula with MGPD margins (ST); an asymmetric logistic copula with lognormal-GPD margins (AL); a Cauchy copula with lognormal margins (CA). Notice that datasets 2G, SN, MO and BL are asymptotically independent, whilst 2T, ST, AL and CA exhibit extreme dependence.
 
Priors were chosen as in Section \ref{sec:priors}. Prior means of $\bm{\mu}_i$ and $\bm{\eta}_i$, $i\in[2]$, were selected around the true values if available, or around values that appeared reasonable after visual investigation of the data histograms, but with large variances. The prior means of the thresholds were  fixed at the 90th empirical quantile.

For all simulations, the codes ran for 25000 iterations, with a burn-in of 5000 and thinning every 20, giving a posterior sample of 1000. Convergence was assessed by looking at trace plots of various functions of the parameters. In all cases, to reduce the number of models to be compared,  the number of gamma mixture components of each marginal was first chosen by fitting different MGPD models. These numbers were then fixed when fitting various mixtures of copulae. Note however that all parameters, both those of the MGPDs and those of the copula densities, were estimated jointly.
 
First notice that, just as for gamma mixtures, only the required copula components have non-zero weights $w_i$ as shown in Table \ref{table1} for the 2G and 2T datasets. Thus, more technical and computationally expensive nonparametric methods are not necessary. The number of mixture components further seems to give an indication of the data asymptotic behaviour: whilst for asymptotically independent datasets all mixtures have the same number of components (first four columns of Table \ref{table1}), asymptotically independent models (Gaussian and Skew-Normal) need a larger number of components than dependent ones (T) for asymptotically dependent simulated data (last four columns of Table \ref{table1}).

\begin{table}[]
\centering
\begin{tabular}{l|cccccccc}
 & 2G & SN & MO & BL & 2T & ST & AL & CA \\
 \hline
G& 2 & 1 & 1 & 1 & 2 & 2 & 1 & 2 \\
Skew-N & 2 & 1 & 1 & 1 & 2 & 2 & 1 & 2 \\
T & 2 & 1 & 1 & 1 & 2 & 1 & 1 & 1
\end{tabular}
\caption{\footnotesize{Number of non-zero copula component weights for Gaussian (G), skew-Normal (skew-N) and Student-T (T) mixtures.}\label{table1}}
\end{table}

The posterior distributions of the degrees of freedom summarized in Table \ref{table2}, being more concentrated around larger values in asymptotically independent datasets, provide a second reliable indicator of the data asymptotic behavior. This is confirmed by the coefficient $\phi$ which takes notably larger values for asymptotically dependent datasets (last line of Table \ref{table2}). The only exception is the dataset from a mixture of T-copulae for which the true number of degrees of freedom is seven: thus a value for $\phi$ around 0.5 is to be expected.   

\begin{table}[]
\centering
\resizebox{\textwidth}{!}{
\begin{tabular}{l|cccccccc}
 & 2G & SN & MO & BL & 2T & ST & AL & CA \\
 \hline
T1 & 3.2 (2.5,4.5) & 28.9 (10.2,135.8) & 38.9 (13.0,154.3) & 13.0 (4.0,157.9) & 2.4 (1.9,3.1) & 5.6 (3.9,9.3) & 7.3 (4.4,16.0) & 0.9 (0.8,1.1) \\
T2 & 16.5 (5.8,141.5) & NA & NA & NA & 9.8 (3.6,51.9) & NA & NA & NA \\
ST1 & 4 (3,6) & 19 (12,29) & 20 (13,29) & 23 (13,32) & 3 (2,3) & 6 (4,12) & 8 (4,21) & 1 (1,1)\\
$\phi$& 0.787 & 0.983 & 0.995 & 0.631 &0.490&0.013&0.191&0
\end{tabular}}
\caption{\footnotesize{Posterior means and 95\% credibility intervals of the degree of freedoms for the Student-T (T1), mixture of two Student-T (T2) and skew-T (ST1) copulae and $\phi$ criterion for the mixture of T-copulae with the most non-zero weights for each dataset.} \label{table2}}
\end{table}

Standard model selection criteria, e.g. BIC \citep{Schwarz1978} and DIC \citep{Spiegelhalter2002}, although giving guidance on the number of mixture components and on the presence of skewness, do not provide information about extreme dependence, possibly because these are mostly influenced by the bulk of the data (see the Supplementary Material).

\section{Applications}

\label{sec:applications}
\begin{figure}
\begin{center}
\begin{subfigure}{0.48\textwidth}
\centering
\includegraphics[scale=0.4]{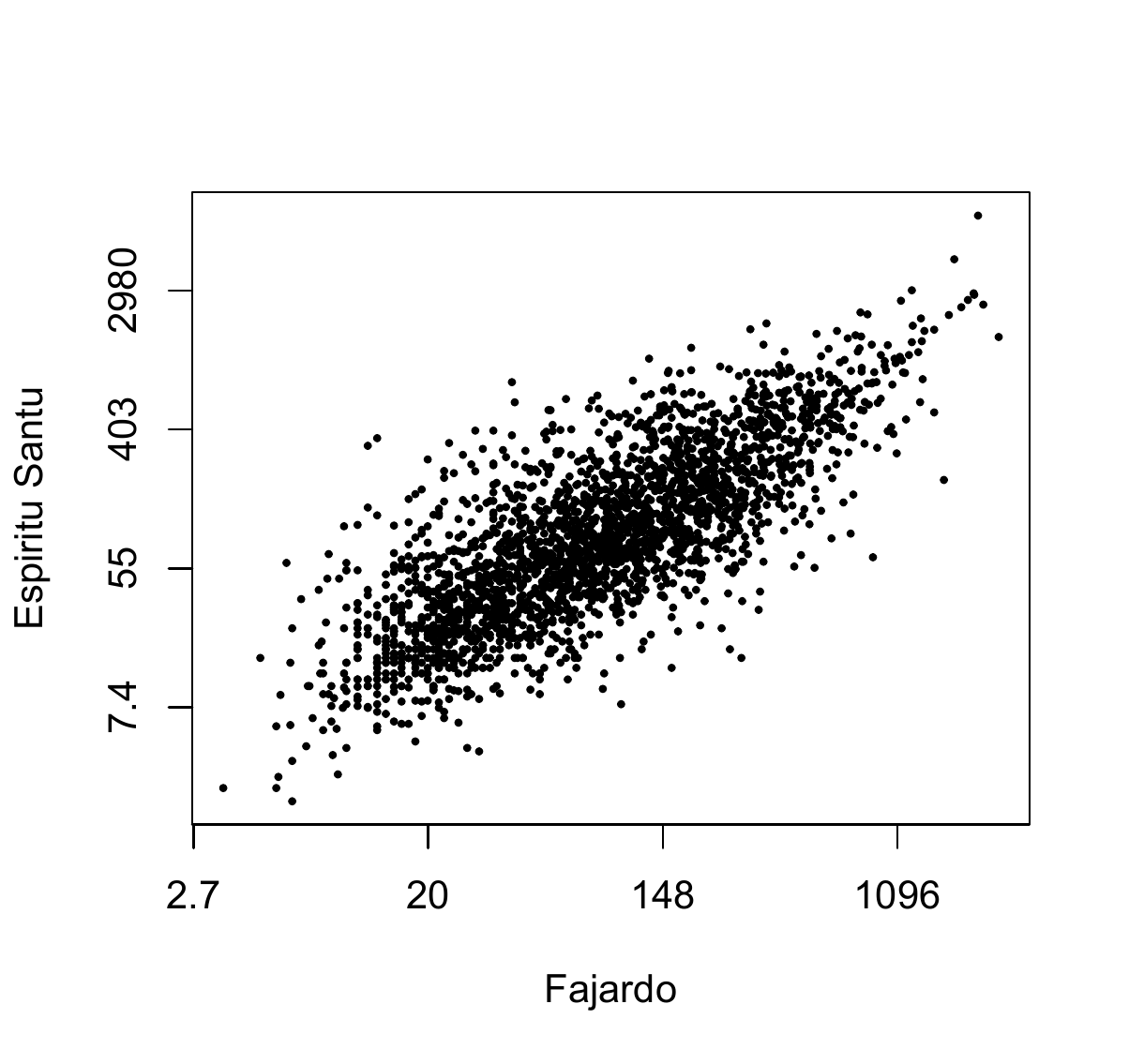}
\vspace{-0.3cm}
\caption{\footnotesize{Puerto Rico rivers' flows (in the log scale, with axes labels on the original scale) }\label{fig:rivers}}
\end{subfigure}
\begin{subfigure}{0.48\textwidth}
\centering
\includegraphics[scale=0.4]{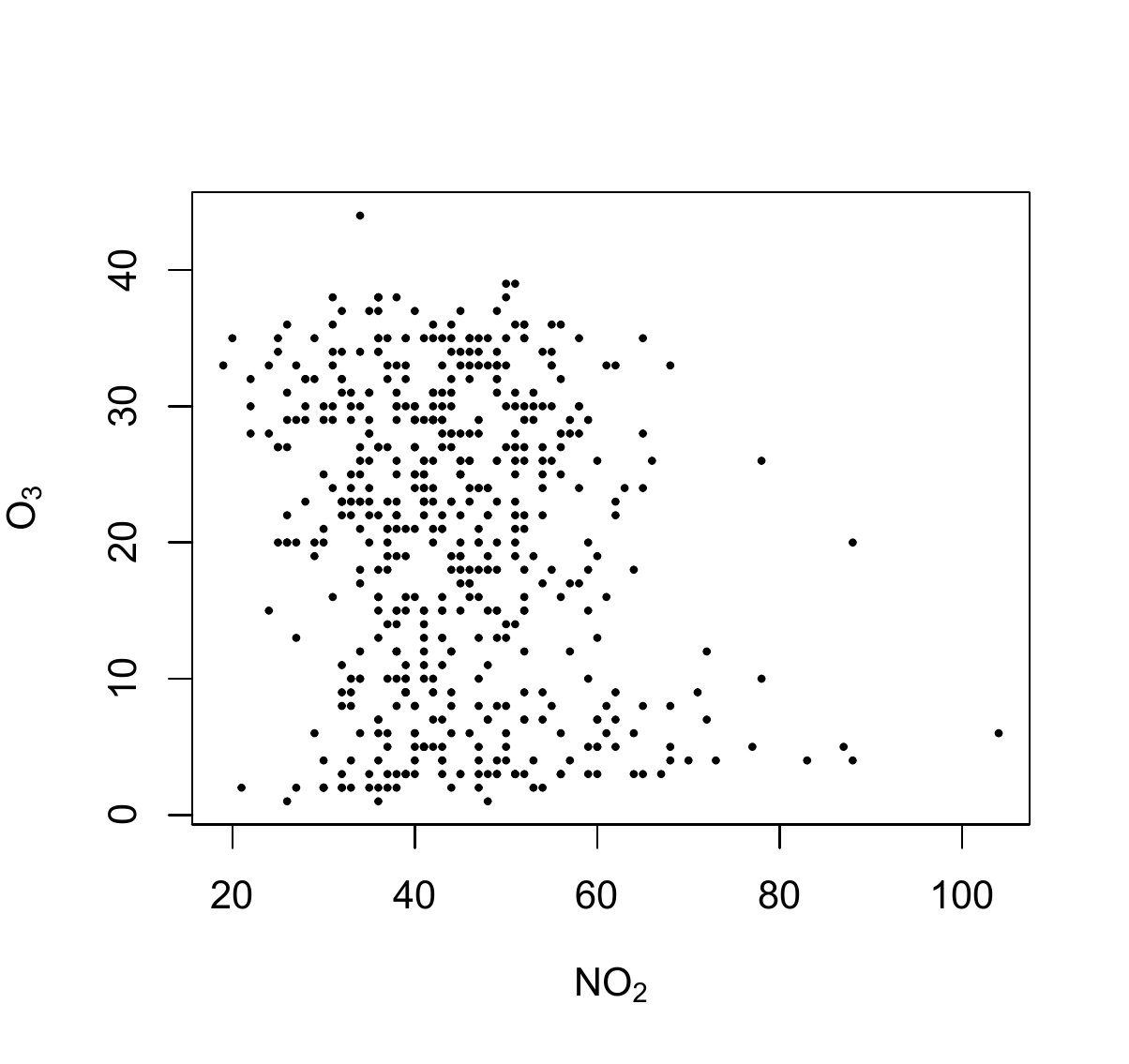}
\vspace{-0.3cm}
\caption{\footnotesize{Leeds pollutants concentrations in the winter months \label{fig:leeds}}}
\end{subfigure}
\end{center}
\vspace{-0.75cm}
\caption{\footnotesize{Datasets from environmental applications.}\label{fig:datasets}}
\end{figure}

Two datasets from environmental applications are analysed next:
\begin{itemize}
\item weekly maxima from August 1966 to June 2016 of the flows of Fajardo and Espiritu Santu rivers in Puerto Rico, comprising 2492 observations  \citep{Nascimento2012}; 
\item  daily maxima of the hourly means during the winter months in 1994-1998 of NO$_2$/O$_3$ concentrations in Leeds, comprising 532 observations  \citep{Heffernan2004}.
\end{itemize}

The Puerto Rico rivers dataset (Figure \ref{fig:rivers}) is freely available at waterdata.usgs.gov, whilst the Leeds pollutants dataset (Figure \ref{fig:leeds}) can be found in R packages. These were chosen for their apparent different asymptotic dependence: in Figure \ref{fig:datasets} the Puerto Rico rivers seem to have strong extreme dependence, whilst the Leeds pollutants appear to have independent extremes \citep[as noted in][]{Heffernan2004}.  In both cases some of the data points were not used for model fitting but to test predictive capabilities of both our and other  approaches. Specifically, 1000 and 100 observations were selected at random and discarded from the Puerto Rico rivers and Leeds contaminants datasets, respectively.

Our approach is compared against the asymptotically independent multivariate Gaussian tail model of \citet{Bortot2000}, the best asymptotically dependent model in the EVD R package \citep{Stephenson2002} and the model of \citet{Ramos2009} that can account for both dependent and independent extremes. For all these models, marginal thresholds were  selected  as in \citet{Ledford1997} at a high empirical quantile of the variable $\min(-\log(\hat{F}_1(X_1))^{-1},-\log(\hat{F}_2(X_2))^{-1})$, where $\hat{F}$ is the empirical df . In this study different empirical quantiles of this variable were used, namely the 90, 95 and 97.5 quantiles\footnote{These values were chosen as they have been used in the  literature \citep{Ledford1997,Ramos2009}.}. For each threshold and marginal,  a GPD was first fitted to the exceedances using a POT approach and then  the data was transformed into Frech\'{e}t margins via empirical df for data below the threshold and GPD df otherwise. Bivariate extreme models  were lastly fitted over the resulting datasets.

\subsection{Model choice}

To start our data analysis the best copula mixture for each dataset is determined. The number of components with non-zero weights suggests that the Puerto Rico rivers dataset might be asymptotically dependent, whilst for the Leeds pollutants datasets extremes appear to be independent. This is because in the latter all mixtures consist of one component only, whilst for the Puerto Rico rivers dataset Gaussian and skew-normal mixtures have two non-zero components. The result of the estimation of the degrees of freedom of the T-copula reported in Table \ref{table:1} confirms this behavior.  Since the posterior credibility intervals of the skewness parameters for all skew-models include zero, we choose the Gaussian for the Leeds contaminants and the T for the Puerto Rico rivers as our favourite mixtures (BIC and DIC values are given in the Supplementary Material).

\begin{table}
\centering
\begin{subtable}{.5\textwidth}
\centering
\begin{tabular}{c|c|c|c}
&Mean & 95\% Int. & $\phi$\\
\hline
Puerto Rico & 5.3 & (3.8,7.9) & 0.003\\
Leeds & 26.2 & (7.7,133.2)&0.93
\end{tabular}
\caption{\footnotesize{Fitting dataset} \label{table:1}}
\end{subtable}
\begin{subtable}{.5\textwidth}
\centering
\begin{tabular}{c|c|c|c}
&Mean & 95\%  Int. & $\phi$\\
\hline
Puerto Rico & 9.89 & (2.70,45.53) & 0.25\\
Leeds & 21.57 & (2.74,107.89)&0.55
\end{tabular}
\caption{\footnotesize{Extreme points only} \label{table:2}}
\end{subtable}
\caption{\footnotesize{Posterior mean and 95\% credibility interval for the degrees of freedom of the T-copula and $\phi$ criterion defined in Section \ref{sec:asce}.\label{tablenew}}}
\end{table}

\begin{figure}
\vspace{-1cm}
\begin{center}
\begin{subfigure}{0.48\textwidth}
\centering
\includegraphics[scale=0.35]{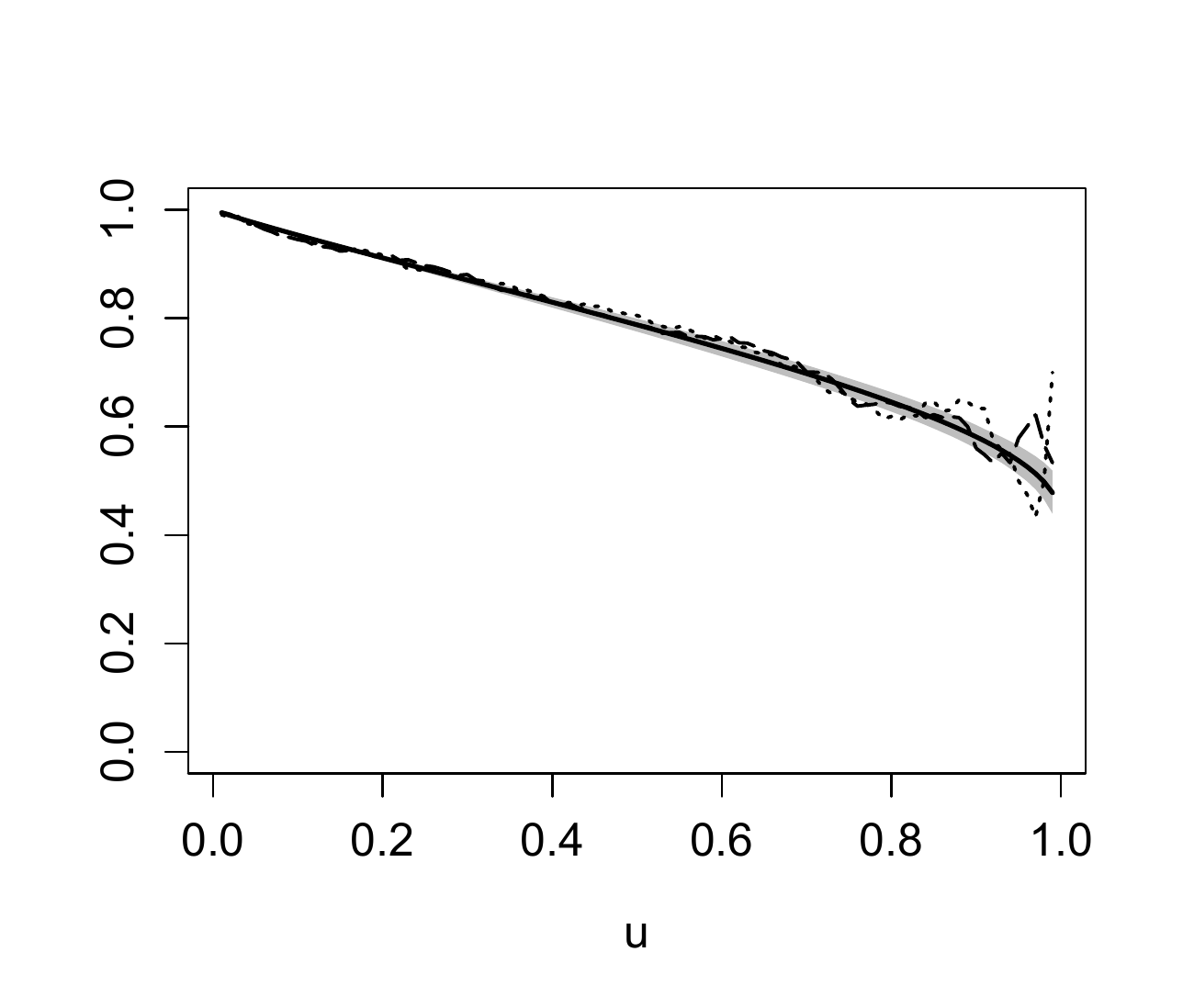}
\vspace{-0.5cm}
\caption{\footnotesize{$\chi(u|\Theta)$: Puerto Rico rivers}\label{fig:deprivers}}
\end{subfigure}
\begin{subfigure}{0.48\textwidth}
\centering
\includegraphics[scale=0.35]{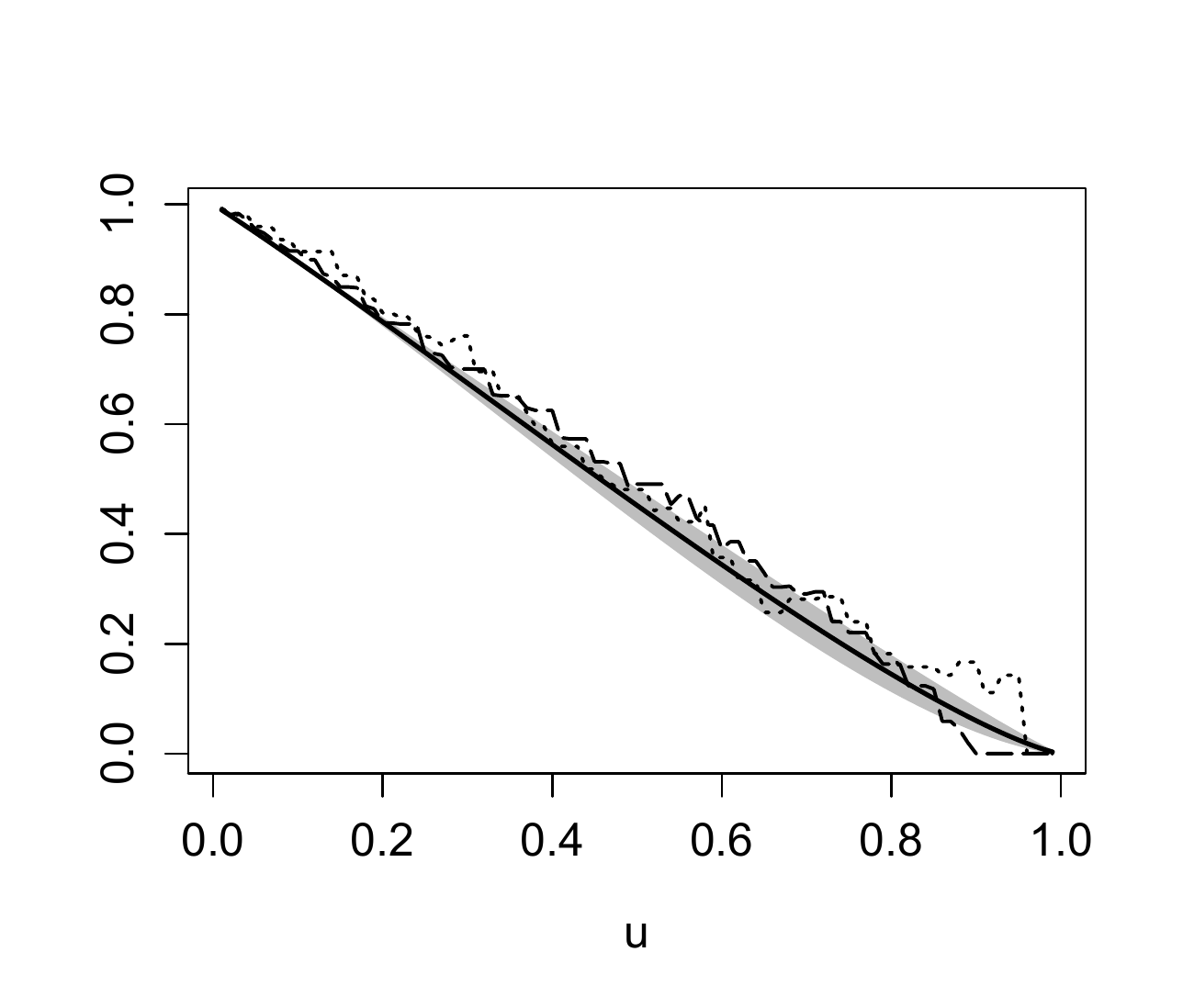}
\vspace{-0.5cm}
\caption{\footnotesize{$\chi(u|\Theta)$: Leeds pollutants}\label{fig:depleeds}}
\end{subfigure}
\end{center}
\vspace{-1.5cm}
\begin{center}
\begin{subfigure}{0.48\textwidth}
\centering
\includegraphics[scale=0.35]{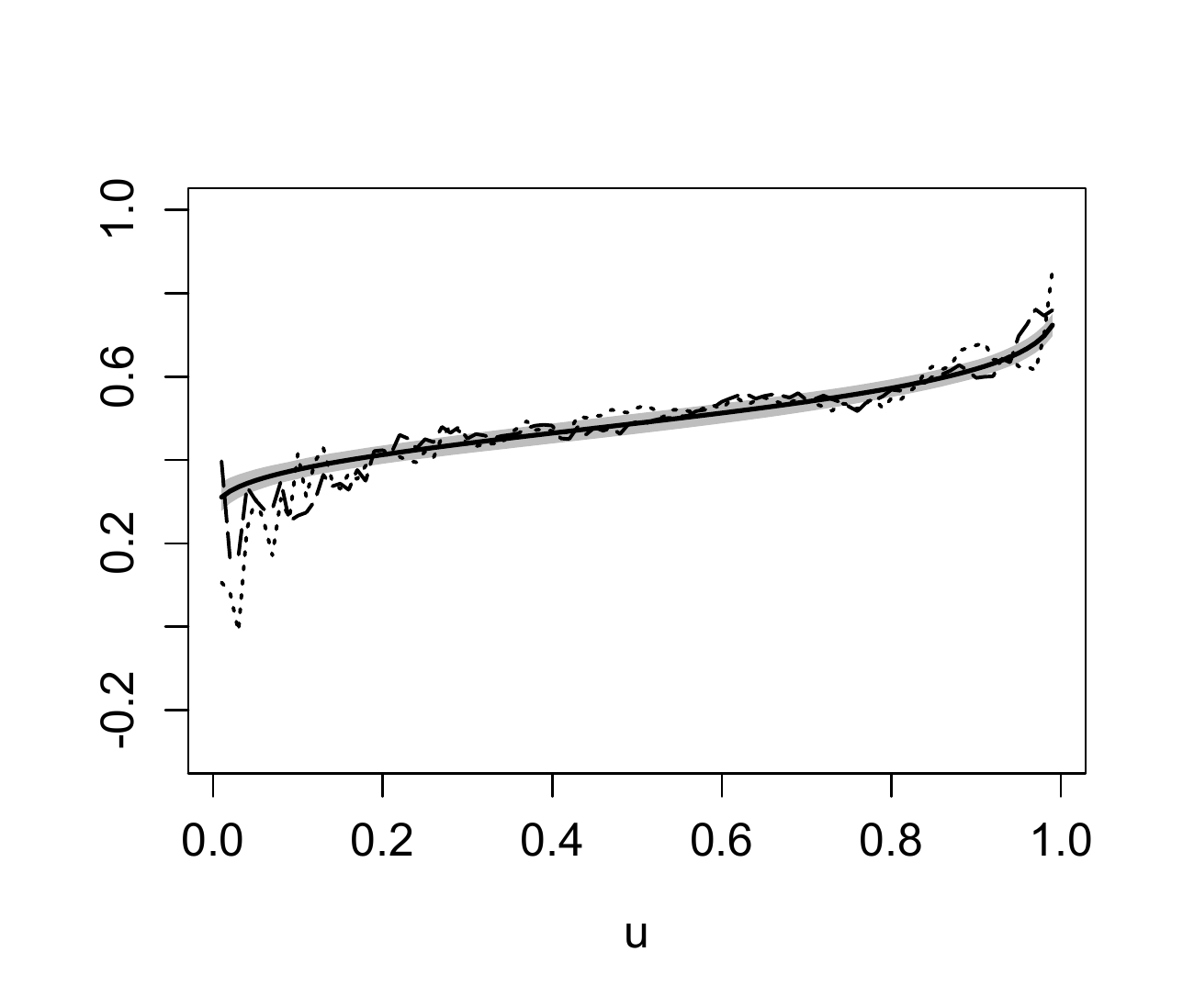}
\vspace{-0.5cm}
\caption{\footnotesize{$\bar{\chi}(u|\Theta)$: Puerto Rico rivers}\label{fig:deprivers2}}
\end{subfigure}
\begin{subfigure}{0.48\textwidth}
\centering
\includegraphics[scale=0.35]{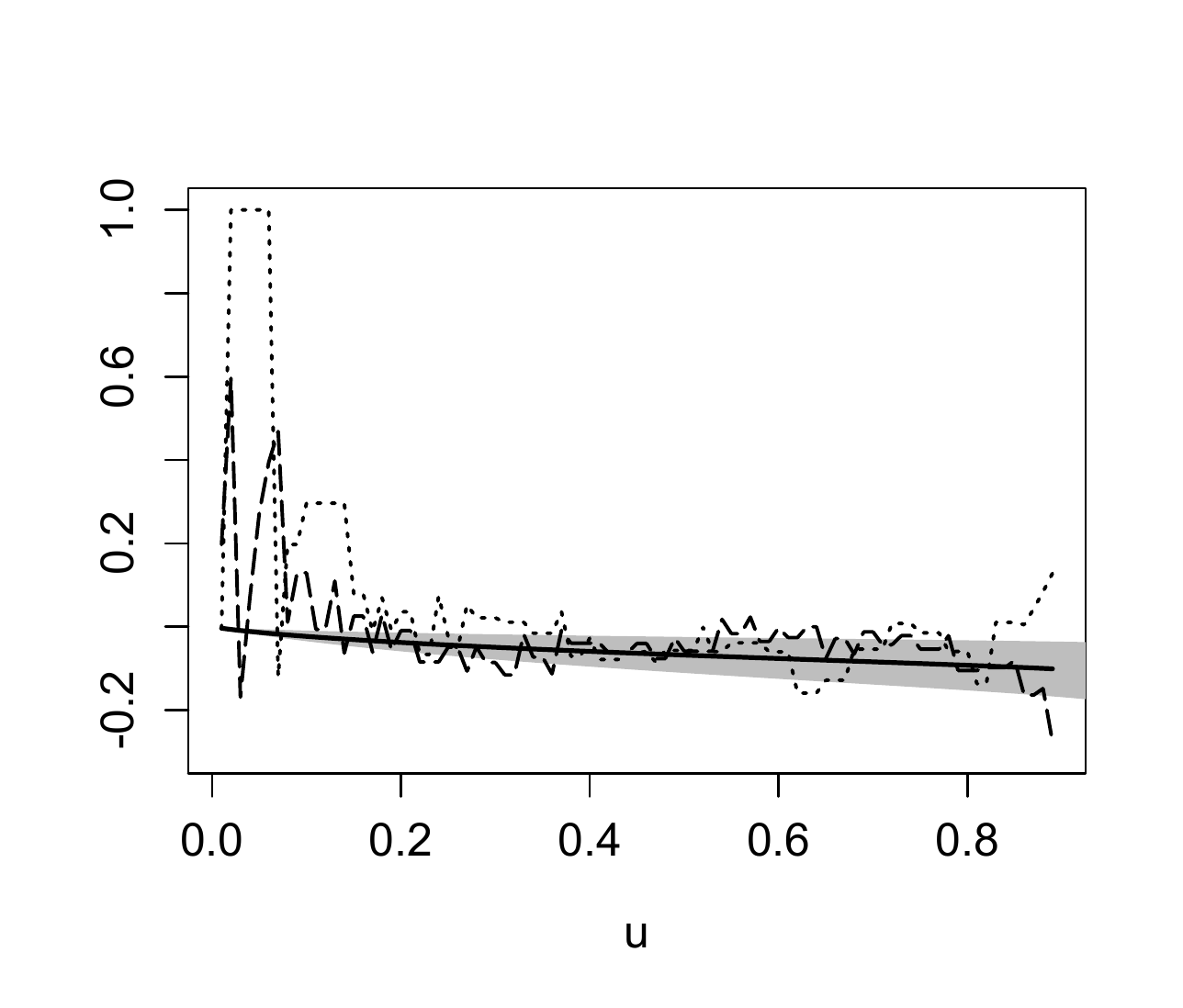}
\vspace{-0.5cm}
\caption{\footnotesize{$\bar{\chi}(u|\Theta)$: Leeds pollutants}\label{fig:depleeds2}}
\end{subfigure}
\end{center}
\vspace{-0.75cm}
\caption{\footnotesize{Posterior estimates of $\chi(u|\Theta)$ and $\bar{\chi}(u|\Theta)$. Full line: posterior mean - Shaded region: 95\% posterior credibility interval - Dashed line: empirical estimate of fitted dataset - Dotted line: empirical estimate of test dataset.}\label{fig:chi}}
\end{figure}

\subsection{Measures of asymptotic dependence}
\label{sec:mes}
In Figures \ref{fig:deprivers} and \ref{fig:depleeds} the posterior estimates of $\chi(u|\Theta)$ for our preferred mixtures are reported. For both applications the posterior means give a good fit to the associated empirical estimates from the fitting and test datasets. These two diagrams give a further indication of asymptotic dependence for the Puerto Rico rivers, as $\chi(u|\Theta)$ tends to $0.5$, and asymptotic independence for Leeds pollutants, as $\chi(u|\Theta)$ goes to zero. Similar conclusions are drawn from the probabilities $\bar{\chi}(u|\Theta)$ in the coefficient of subasymptotic dependence reported in Figures \ref{fig:deprivers2} and \ref{fig:depleeds2}. To the limit these confirm the asymptotic behaviour shown by $\chi(u|\Theta)$, since for instance for the Puerto Rico rivers $\bar{\chi}(u|\Theta)$ goes to one. 

\subsection{Predictions}

\begin{table}
\centering
\resizebox{\textwidth}{!}{
\begin{tabular}{c|c|c|c|c|c|c}
&Empirical & Marginal & Joint & POT 90 & POT 95 & POT 97.5 \\
\hline
Fajardo  & [1710,1800]& 1900 (1554,2544) & 1865 (1564,2289) & 1875& 1975& 2031\\
Espiritu Santu  & [1350,1380] & 1463 (1215,1886) &1388 (1210,1663) &  1464 & 1459 & 1477
\end{tabular}}
\caption{\footnotesize{Posterior summaries of $q(p|\Theta)$ for the Fajardo and Espiritu Santu rivers with $p=0.005$: Empirical - empirical quantile from test dataset; Marginal - estimated quantiles using a marginal MGPD model; Joint - estimated quantiles using the bivariate approach; POT - estimated quantiles using a POT approach at different thresholds.}\label{tablenew1}}
\end{table}

\begin{table}
\begin{center}
\scalebox{0.7}{
\begin{tabular}{c|ccc}
\multicolumn{4}{c}{\hspace{0.4cm}\textbf{Puerto Rico rivers}}\\
&\multicolumn{3}{c}{$(x_1,x_2)$}\\
Models&  (720,730) & (900,780) & (1300,1100)\\
\hline
Emp. Pred. & 0.015 & 0.010 & 0.005\\
\hline
T1&0.0175 &0.0115 &0.0044 \\
95\% CI &(0.0138,0.0220)&(0.0086,0.0149)&(0.0028,0.0069)\\
\hline
EVD 90 & 0.0209 &0.0141&0.0057 \\
EVD 95 &  0.0214 & 0.0145 &0.0058 \\
EVD 97.5  &0.0211& 0.0154 & 0.0064 \\
Bortot 90& 0.0186 &0.0122&0.0046 \\
Bortot 95 &0.0205 & 0.0135 &0.0050 \\
Bortot 97.5  &0.0216 & 0.0153 & 0.0060 \\
Ramos 90 & 0.0203 &0.0135& 0.0054 \\
Ramos 95 & 0.0201 & 0.0136 & 0.0054 \\
Ramos 97.5  &0.0207 & 0.0149 & 0.0062 
\end{tabular}}
\,\,\,\,\,\,\,\,\,\,\,\,\,
\scalebox{0.7}{
\begin{tabular}{c|cc}
\multicolumn{3}{c}{\textbf{Leeds pollutants}}\\
&\multicolumn{2}{c}{$(x_1,x_2)$}\\
&   (55,32) & (58,33)\\
\hline
Emp. Pred. & 0.020 & 0.010\\
\hline
G1&0.0188 &0.0104 \\
95\% CI &(0.0126,0.0265)&(0.0065,0.0118)\\
\hline
EVD 90 & 0.0549&0.0405 \\
EVD 95 &   0.0854 &0.0607\\
EVD 97.5   & 0.0875 & 0.0635 \\
Bortot 90& 0.0161&0.0085 \\
Bortot 95 &0.0133 &0.071\\
Bortot 97.5  & 0.0099 & 0.0050 \\
Ramos 90  &0.0114& 0.0052 \\
Ramos 95 & 0.0122 & 0.0049 \\
Ramos 97.5   & 0.0093 & 0.0034 
\end{tabular}}
\end{center}
\vspace{-0.5cm}
\caption{\footnotesize{Posterior summaries of $E(x_1,x_2|\Theta)$ for various $(x_1,x_2)$ and estimates from competing models.}\label{table:table}}
\end{table}
\begin{figure}
\begin{center}
\begin{subfigure}{0.19\textwidth}
\centering
\includegraphics[scale=0.3]{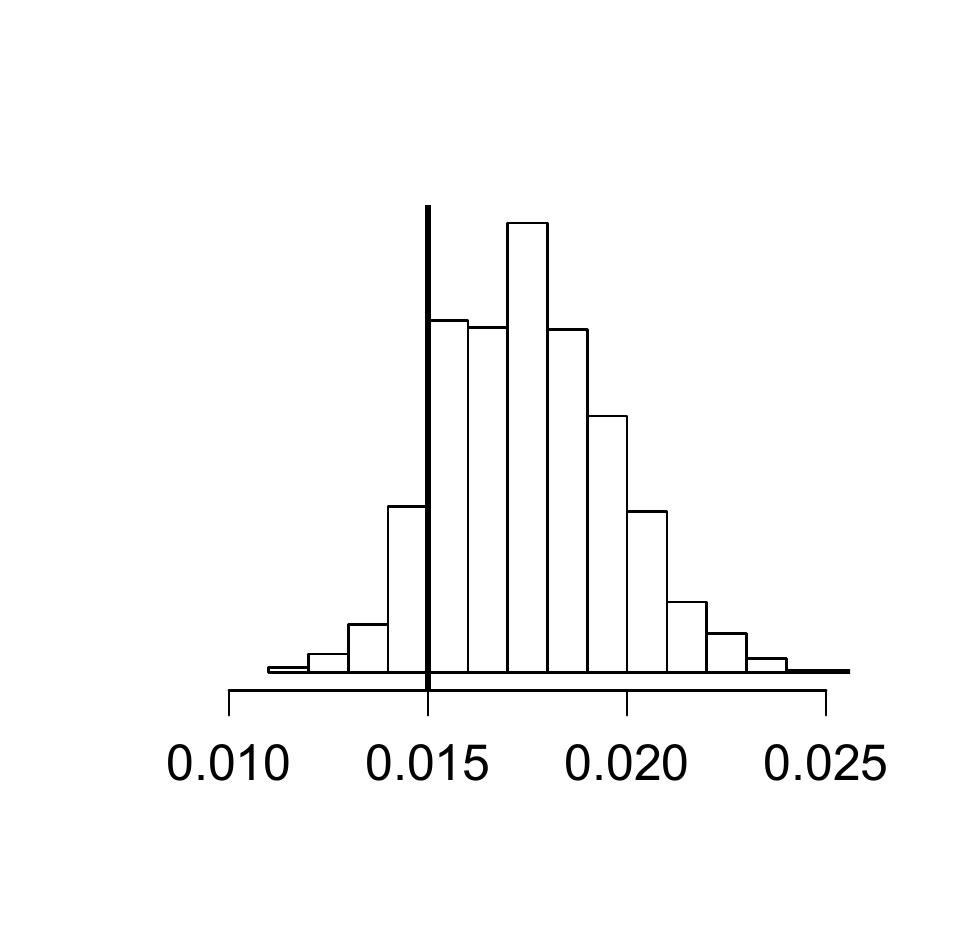}
\vspace{-0.3cm}
\caption{\footnotesize{(720,730)}}
\end{subfigure}
\begin{subfigure}{0.19\textwidth}
\centering
\includegraphics[scale=0.3]{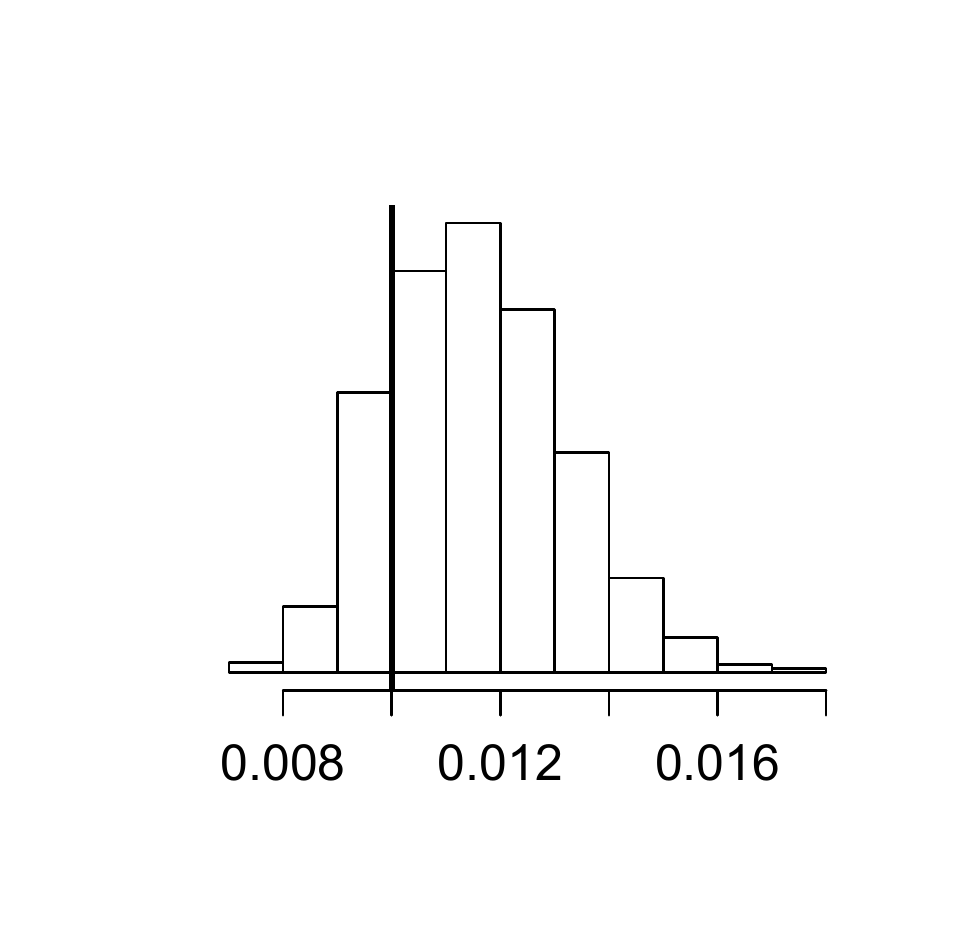}
\vspace{-0.3cm}
\caption{\footnotesize{(900,780)}}
\end{subfigure}
\begin{subfigure}{0.19\textwidth}
\centering
\includegraphics[scale=0.3]{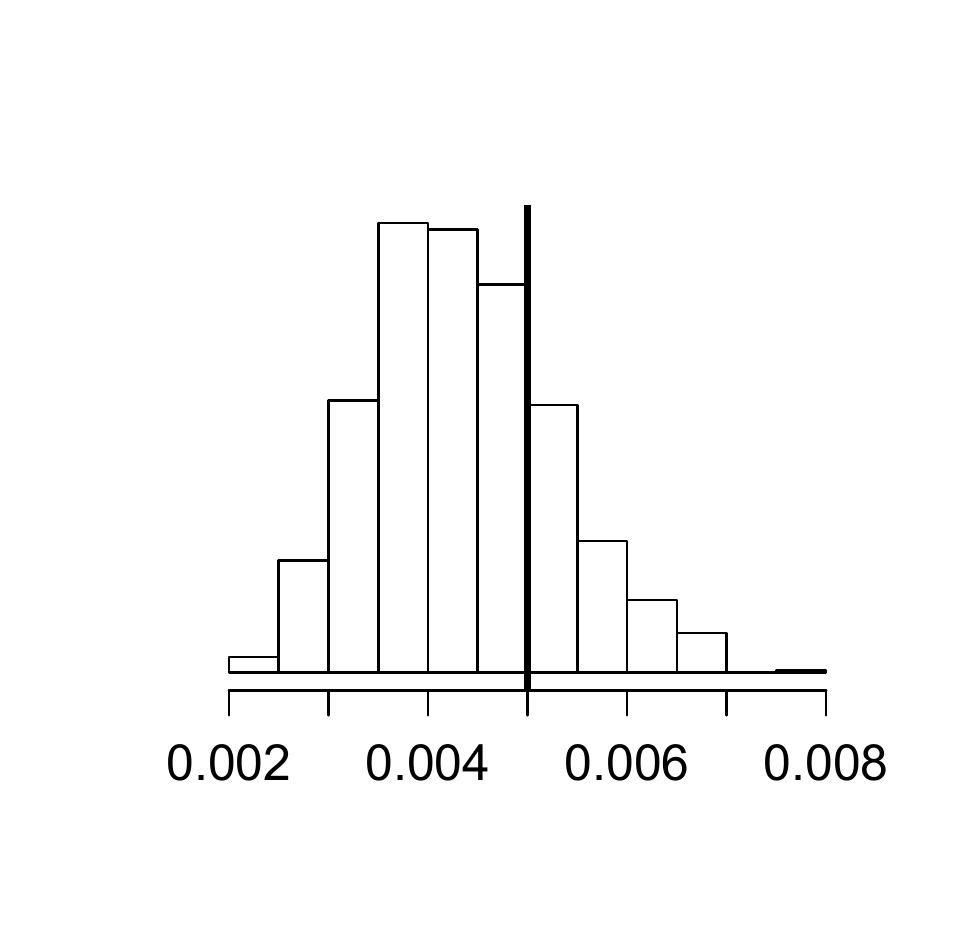}
\vspace{-0.3cm}
\caption{\footnotesize{(1300,1100)}}
\end{subfigure}
\begin{subfigure}{0.19\textwidth}
\centering
\includegraphics[scale=0.3]{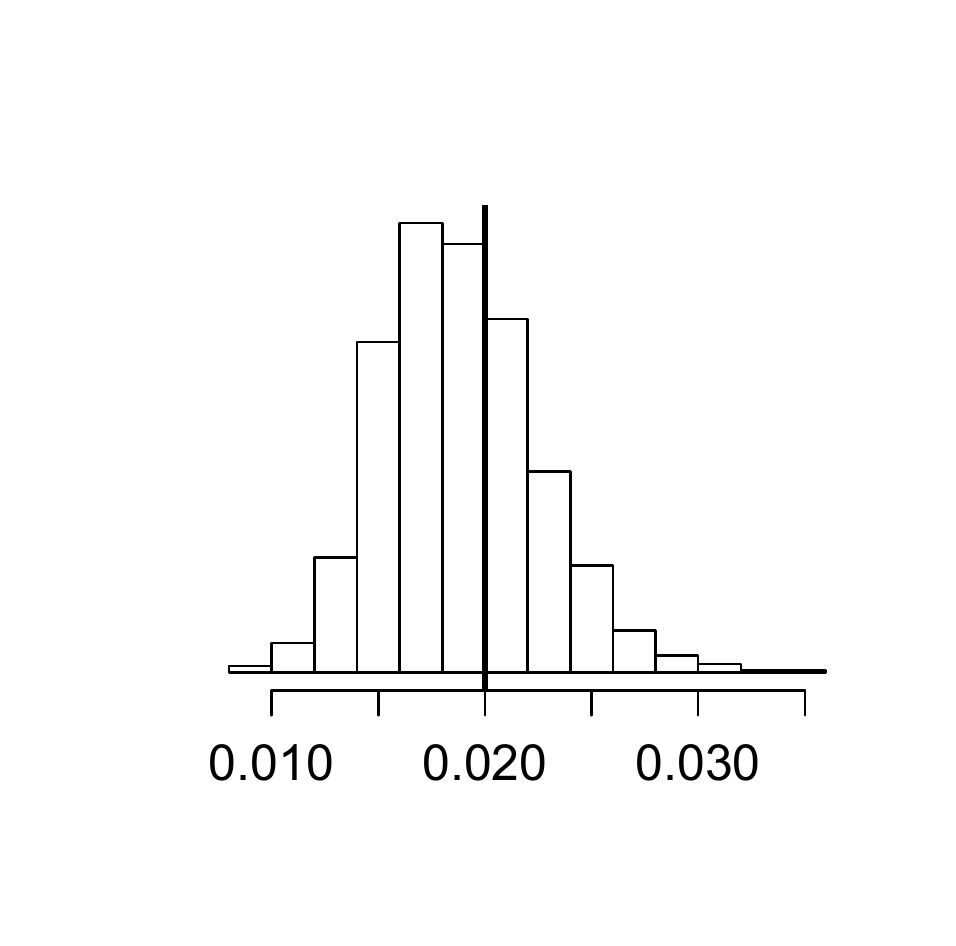}
\vspace{-0.3cm}
\caption{\footnotesize{(55,32)}}
\end{subfigure}
\begin{subfigure}{0.19\textwidth}
\centering
\includegraphics[scale=0.3]{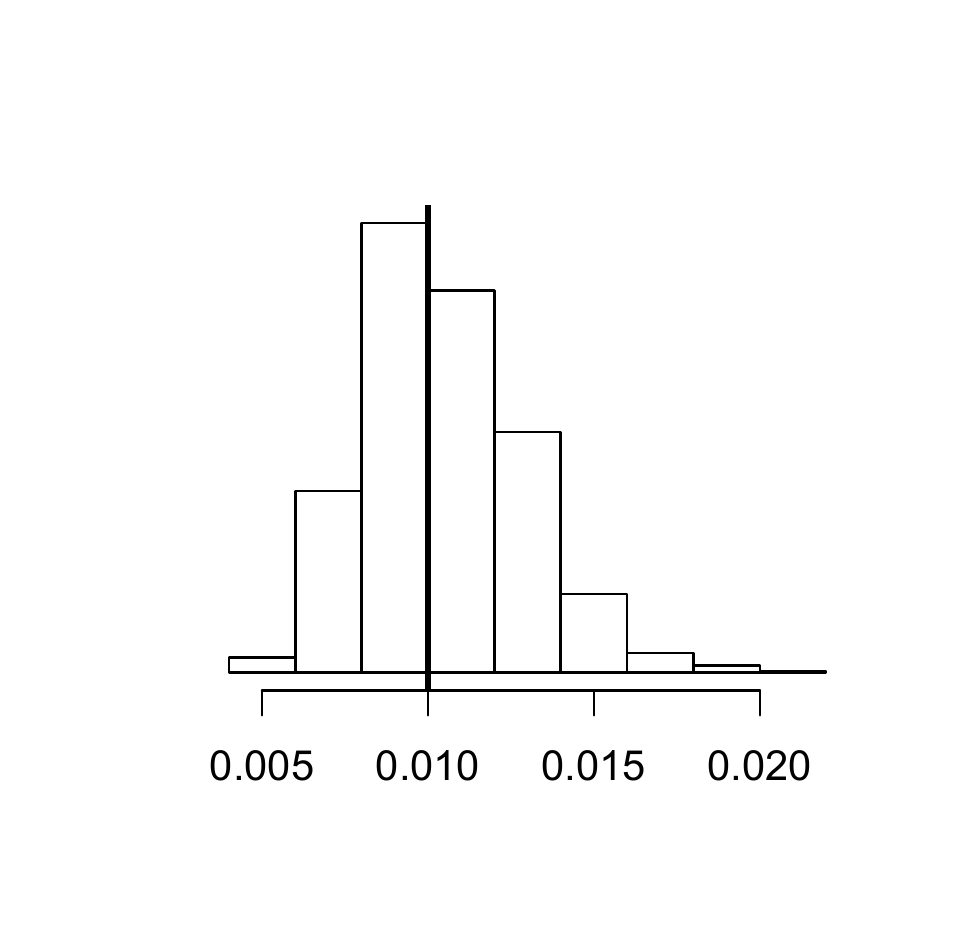}
\vspace{-0.3cm}
\caption{\footnotesize{(58,33)}}
\end{subfigure}
\end{center}
\vspace{-0.75cm}
\caption{\footnotesize{ Posterior distribution of $E(x_1,x_2|\Theta)$ for various $(x_1,x_2)$ - Figures (a)-(c): Puerto Rico rivers - Figures (d)-(e): Leeds pollutants - Vertical lines: empirical predictive estimates.}\label{fig:hist}}
\end{figure}

The performance in extreme predictions of our approach is studied next. Marginally, as already noted in \citet{Nascimento2012}, the MGPD can outperform the POT methodology. This is reported in Table \ref{tablenew1} for the Puerto Rico rivers. Importantly, the table shows that joint modelling gives not only much narrower posterior credibility intervals than a simpler MGPD model, but also predicted values closer to the empirical ones.

The properties of the posterior distributions of $E(x_1,x_2|\Theta)$ for various pairs $(x_1,x_2)$ whose elements exceed the used thresholds are summarized in Table \ref{table:table} together with estimates from the other approaches considered as well as the empirical probabilities of the test data.  Our approach outperforms competing ones for the Leeds pollutant dataset in all pairs. For the Puerto Rico rivers dataset, our estimates are more accurate for all pairs but the one associated to an exceedance probability of $0.005$. In all cases, the 95\% posterior credibility intervals from our mixtures include the empirical probability.  In Figure \ref{fig:hist} is further reported the posterior distributions of  $E(x_1,x_2|\Theta)$ for the pairs considered in Table \ref{table:table}: these are in general not available using the approaches reviewed in Section \ref{sec:multi}.

\begin{figure}
\begin{center}
\begin{subfigure}{0.48\textwidth}
\centering
\includegraphics[scale=0.4]{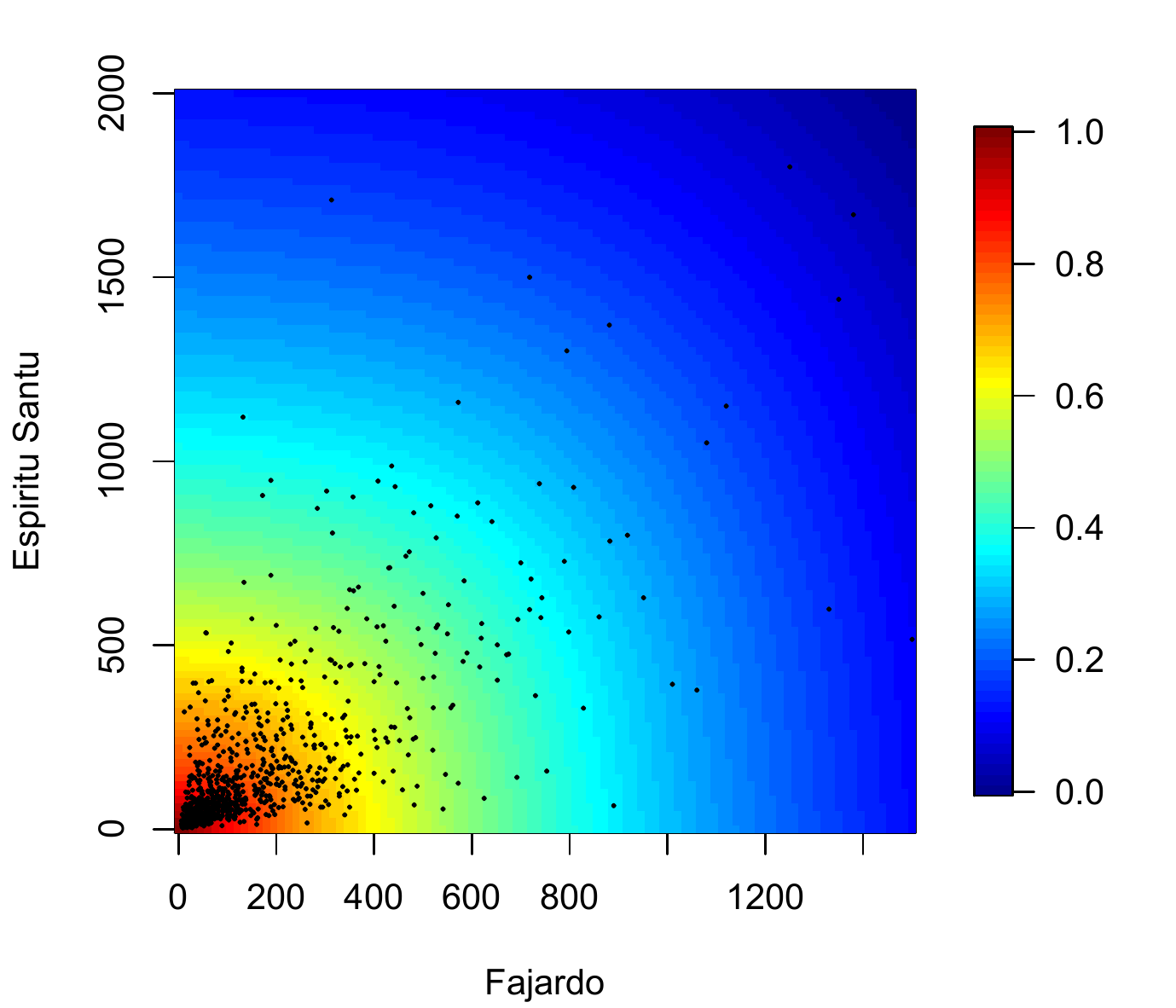}
\vspace{-0.3cm}
\caption{\footnotesize{Puerto Rico rivers dataset}\label{fig:deprivers1}}
\end{subfigure}
\begin{subfigure}{0.48\textwidth}
\centering
\includegraphics[scale=0.4]{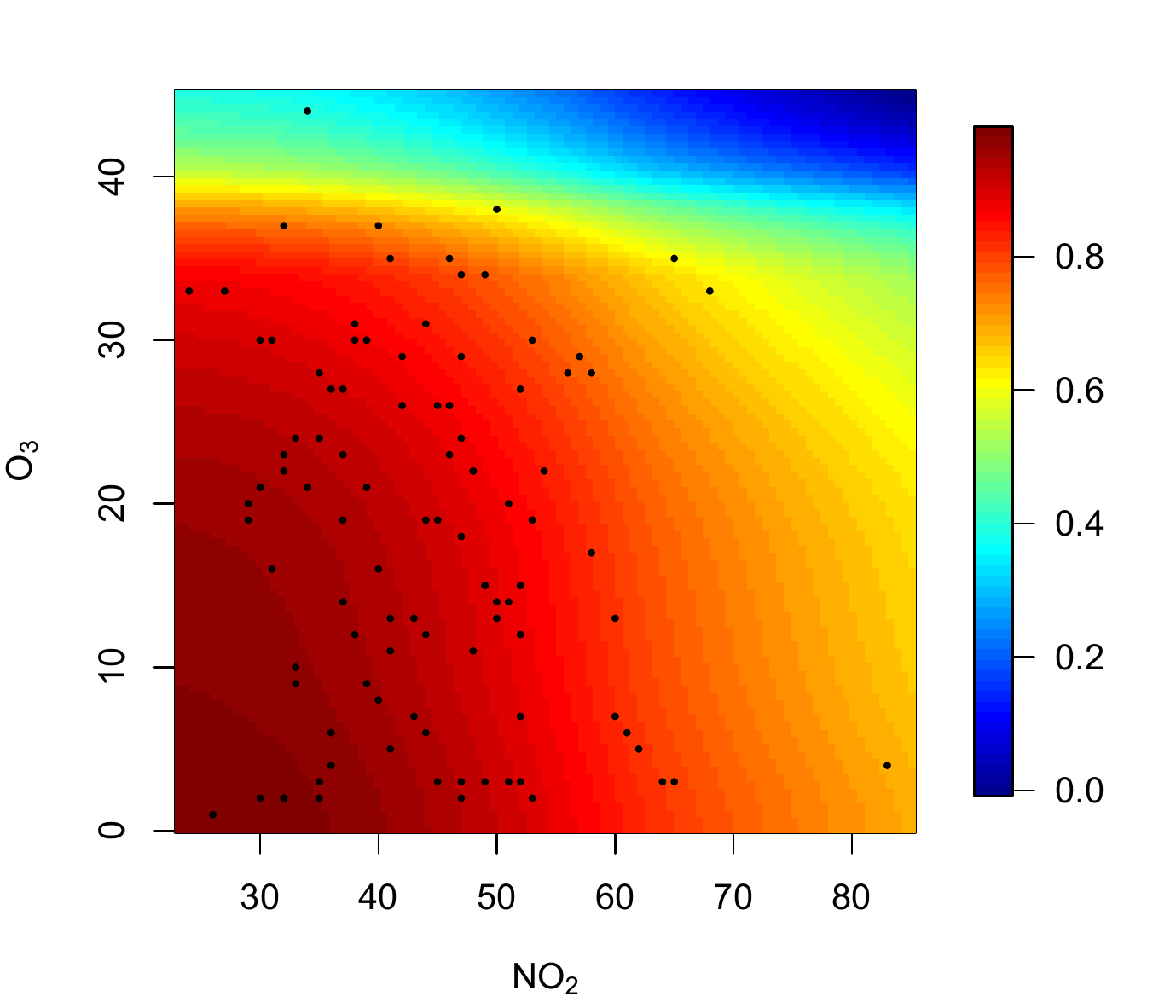}
\vspace{-0.3cm}
\caption{\footnotesize{Leeds pollutants dataset}\label{fig:depleeds1}}
\end{subfigure}
\end{center}
\vspace{-0.75cm}
\caption{\footnotesize{Predictive probabilities of joint exceedance together with predictive datasets.}\label{fig:map}}
\end{figure}

Lastly, Figure \ref{fig:map} reports the Monte Carlo estimates of the predictive probabilities of exceedance  $ E(\bm{x}_{m+1}|\bm{x})$.  Each point $(x_1,x_2)$ of this map gives the probability of a future observation that is larger than both $x_1$ and $x_2$. These provide an intuitive description of the overall behavior of the test datasets. Again, such predictive summaries are often not available for other approaches.

\subsection{Effect of the bulk on estimation of extreme dependence}
\label{sec:uff}

An analysis over a subset of the full datasets, including only points considered extreme, is next carried out to ascertain whether the bulk of the data affects our tail estimation approach. The extreme points are selected as follows: first only observations that exceed the chosen thresholds in both marginals are retained (as in Figure \ref{fig:upper}); for the Puerto Rico rivers application the threshold locations are chosen at the posterior means of the thresholds of the T-copula model (giving 190 observations); for the Leeds pollutants the thresholds were selected to give a marginal probability of exceedance of 0.3 as in \citet{Heffernan2004} (giving 49 observations); lastly the margins of the resulting data points are transformed to the uniform scale via the empirical df.

Mixtures of T-copulae are first fitted to these datasets to investigate whether the asymptotic dependence behaviour chosen by looking at the full dataset is confirmed when considering only extreme points. The results of this analysis summarized in Table \ref{table:2} confirm the asymptotic behaviors identified in Section \ref{sec:mes}, but give much larger posterior credibility intervals to the degrees of freedom and thus uncertainty about the true extreme regime.

Having assessed the asymptotic dependence structure over the extreme points only, the extreme-value copulae \citep{Gudendorf2010} associated to the T and Gaussian copulae are fitted to the extreme datasets of the Puerto Rico rivers and Leeds pollutants applications, respectively. However, for the Puerto Rico rivers a Gumbel copula  given by
\[
G(v_1,v_2)=\exp \!\left[-\left((-\log(v_1))^{\theta }+(-\log(v_2))^{\theta }\right)^{1/\theta }\right], \hspace{0.5cm} v_1,v_2\in[0,1], \theta\in [1,+\infty)
\]
 is used instead since this has an almost identical Pickands dependence function to the one of the extreme T-copula \citep{Demarta2005}. For the Leeds pollutants a Gaussian copula is used since the associated extreme copula would simply be an independent one. Table \ref{unaltra} summarizes the posterior distributions of the relevant coefficients of dependence when estimated using the full dataset or the extreme points only. In both cases the posterior means are around the values of the empirical coefficients reported in Figure \ref{fig:chi}, but importantly the credibility intervals are narrower for the full dataset. 

\begin{table}
\begin{center}
\begin{tabular}{c|c}
&Puerto Rico rivers: $\chi$\\
\hline
Full dataset& 0.45 (0.39,0.50)\\
Extreme points & 0.43 (0.35,0.51)
\end{tabular}
\,\,\,\,\,\,\,\,
\begin{tabular}{c|c}
&Leeds pollutants: $\bar{\chi}$\\
\hline
Full dataset & -0.13 (-0.21,-0.04)\\
Extreme points & -0.23 (-0.48,0.08)
\end{tabular}
\end{center}
\caption{\footnotesize{Posterior means and 95\% credibility intervals for the coefficients of asymptotic dependence (Puerto Rico rivers) and sub-asymptotic dependence (Leeds pollutants)\label{unaltra}.}}
\end{table}

\section{Discussion}
\label{sec:discussion}
In this work a new flexible approach for the estimation and prediction of extremes and joint exceedances was introduced. The issue of model choice between the various mixtures was investigated as well as the performance of our approach in extremes' predictions. The results suggest that our Bayesian semiparametric approach outperformed  other bivariate approaches in predicting new extreme observations for the applications considered, whilst also allowing for the study of not only extreme but also overall dependence structures.  Furthermore, great attention was devoted to the identification of the extreme dependence behaviour by defining the new criterion $\phi$ which gives a probabilistic judgement on the possibility of asymptotic dependence. 

A natural extension of the approach described here could consider two different copulae specification in disjoint subsets of $\mathbb{R}^2_+$. Such subsets might correspond to the ones defined by the thresholds illustrated in Figure \ref{fig:thresholds}.  Such distinction would allow for the use of the the full dataset whilst specifying a different dependence pattern for the extreme region, should one wishes to do so. So for instance the likelihood could be defined as 
\[
f(x_1,x_2)=\left\{
\begin{array}{ll}
c_b(F_1(x_1),F_2(x_2))f_1(x_1)f_2(x_2), & \mbox{if } (x_1,x_2)\in B,\\
Kc_t(F_1(x_1),F_2(x_2))f_1(x_1)f_2(x_2), & \mbox{otherwise, }
\end{array}
\right.
\]
where $c_b$ and $c_t $ are two different copula densities, $K$ is a normalizing constant and $B\subset\mathbb{R}^2$ is the region including non-extreme points. This more general specification brings in extra components and complications ($K$ depends on model parameters in a non-trivial form) and handling them is not so straightforward. Solutions for these issues  are the subject of ongoing research.

Although in this paper the focus was mainly on bivariate problems,  multivariate extensions are readily available. For instance, mixtures of $d$-variate elliptical copulae could be considered. A full definition of the approach would then be completed by an appropriate prior for the covariance matrix, for instance an inverse-Wishart, and an appropriate identification constraint for matrices, for example based on the determinant.

But more interestingly, since different pairs of variables could be defined to have a different asymptotic dependence, the overall density could be defined via vine-copulae \citep{Bedford2002}. For instance, in the trivariate case the overall density via a vine-copula decomposition can be written as
\[
f(x_1,x_2,x_3)=c_{12}(F_1(x_1),F_2(x_2))c_{23}(F_2(x_2),F_3(x_3))c_{13|2}(F_{1|2}(x_1|x_2),F_{3|2}(x_3|x_2))\prod_{i\in[3]}f_i(x_i)
\]
where $F_{i|j}(x_i|x_j)=\partial C_{ij}(F_i(x_i),F_j(x_j))/\partial F_j(x_j)$ and the $c$'s are bivariate copula densities. The investigation of such models will be the focus of future research.

\section*{Acknowledgements}
The authors gratefully acknowledge CAPES and CNPq for financial support. Most of this work was carried out at the Instituto de Matem\'{a}tica at the Universidade Federal do Rio de Janeiro whilst ML held a CAPES postdoctoral fellowship. The authors gratefully thank Jonathan Tawn and Miguel de Carvalho for insightful  comments on previous versions of the manuscript.

\bibliographystyle{Chicago}

\bibliography{Bibliography}

\appendix
\section{Copula densities}
For all the copulae below we let $F_i(x_i|\Theta_i)$, $i\in[2]$, be the df of an MGPD.

\label{appendixA}

\noindent\textbf{Gaussian copula}\\
 In the bivariate case the Gaussian copula density depends on a correlation parameter $\rho\in[-1,1]$ and can be written as
\begin{equation*}
c(F_1(x_1|\Theta_1),F_2(x_2|\Theta_2)|\rho_i)=\frac{1}{\sqrt{1-\rho_i^2}}\exp\left(\frac{2\rho_iz_1z_2-\rho_i^2(z_1^2+z_2^2)}{2(1-\rho_i^2)}\right),
\end{equation*}
where $z_i=\Phi^{-1}(F_i(x_i|\Theta_i))$ and $\Phi$ is the standard univariate normal df.

\noindent\textbf{T-copula}\\
 In the bivariate case the T-copula density depends on a correlation parameter $\rho\in[-1,1]$ and degrees of freedom $v\in\mathbb{R}_+$ and can be written as
\begin{equation*}
c(x_1,x_2|\rho,v)=\frac{\Gamma\left(\frac{v}{2}\right)}{\sqrt{1-\rho_i^2}}\frac{\Gamma\left(\frac{v+2}{2}\right)}{\Gamma\left(\frac{v+1}{2}\right)^2}\frac{\left(1+\frac{z_1^2}{v}+\frac{z_2^2}{v}+\frac{z_1^2z_2^2}{v^2}\right)^{(v+1)/2}}{\left(1+\frac{z_1^2+z_2^2-2\rho_iz_1z_2}{v(1-\rho_i^2)}\right)^{(v+2)/2}},
\end{equation*}
where $z_i=T_v^{-1}(F_i(x_i|\Theta_i))$ and $T_v$ is the standard univariate T df with $v$ degrees of freedom.

\noindent\textbf{Skew-Normal copula}\\
For $\rho\in[-1,1]$ and $\delta_i\in(-1,1)$, $i\in[2]$, define
\begin{equation}
\begin{array}{cc}
\lambda_i=\delta_i/\sqrt{1-\delta_i^2},&
\psi=\rho\sqrt{1-\delta_1}\sqrt{1-\delta_2}+\delta_1\delta_2,\\
\alpha_1=\frac{\delta_1-\delta_2\psi}{((1-\psi^2)(1-\psi^2-\delta_1^2-\delta_2^2+2\psi\delta_1\delta_2))^{1/2}},&
\alpha_2=\frac{\delta_2-\delta_1\psi}{((1-\psi^2)(1-\psi^2-\delta_1^2-\delta_2^2+2\psi\delta_1\delta_2))^{1/2}}.
\end{array}
\label{eq:skew3}
\end{equation}
The density of a bivariate skew-normal copula then depends on the parameters defined in equation (\ref{eq:skew3}) and can be written as
\[
c(x_1,x_2|\psi,\alpha_{1},\alpha_{2},\lambda_1,\lambda_2)=\frac{sn(z_1,z_2|\psi,\alpha_{1},\alpha_{2})}{sn(z_1|\lambda_1)sn(z_2|\lambda_2)},
\]
where $sn(z|\lambda)=2\phi(z)\Phi(\lambda z)$ - with $\phi$ the density of a standard normal distribution - $z_i=SN^{-1}(F_i(x_i|\Theta_i)|\lambda_i)$ - with $SN$ the df associated to the density $sn$ - and $sn(z_1,z_2|\psi,\alpha_{1},\alpha_{2})=2\phi_{\psi}(x_1,x_2|\psi)\Phi(\alpha_1x_1+\alpha_2x_2)$ - with $\phi_{\psi}$ the density of a bivariate standard normal distribution with correlation $\psi$.
 
\noindent\textbf{Skew-T copula}\\
For $\rho\in[-1,1]$, $v\in\mathbb{R}$ and $\delta_i\in(-1,1)$, $i\in[2]$, define $\lambda_i$, $\psi$ and $\alpha_i$ as in equation (\ref{eq:skew3}). The density of a bivariate skew-T copula can be written as 
\begin{equation*}
c(x_1,x_2|\psi,v,\alpha_1,\alpha_2,\lambda_1,\lambda_2)=\frac{st(z_1,z_2|\psi,v,\alpha_1,\alpha_2)}{st(z_1|\lambda_1,v)st(z_2|\lambda_2,v)},
\end{equation*}
where $st(z|\lambda,v)=2 t_v(z|v)T_v(\lambda z\sqrt{(v+1)/(z^2+v)})$ - with $t_v$ the density of a standard univariate T with $v$ degrees of freedom - $z_i=ST^{-1}(F_i(x_i|\Theta_i)|\lambda_i,v)$ - with $ST$ the df associated to the density $st$ - and 
\[
st(x_1,x_2|\cdot)=2t_{\psi,v}(x_1,x_2|\psi,v)T_{v+2}\left(\frac{\alpha_1 x_1+\alpha_2 x_2}{\sqrt{(x_1^2+x_2^2-2\psi x_1x_2+v(1-\psi^2))/((v+2)(1-\psi^2))}}\right)
\]
- with $t_{\psi,v}$ the density of a bivariate standard T distribution with $v$ degrees of freedom and correlation $\psi$.

\section{MCMC algorithm}
\label{appendixB}
Sampling is carried out in blocks with Metropolis-Hastings proposals. At each iteration we first sample the copula parameters and copula mixture weights, and then the marginal parameters for each of the marginals. For the marginals we use the steps outlined in \citet{Nascimento2012} and therefore we do not report them here. However, in our case the acceptance probabilities are computed with respect to the posterior in equation (\ref{eq:posterior}).

At iteration $s$ parameters are updated as follows.

\begin{itemize}
\item Sampling $\rho_i$, for $i\in[k]$. 

Since the correlation $\rho_i\in[-1,1]$, the proposal kernel is taken as the truncated Normal distribution
$
\rho_i^*|\rho_i^{(s)}\sim\mathcal{N}(\rho_i^{}(s),V_{\rho_i})\mathbbm{1}_{-1\leq \rho_1^{(s+1)}<\cdots<\rho_{i-1}^{(s+1)}<\rho_i^{(s)}<\cdots<\rho_n^{(s)}\leq 1},
$   
where $\rho_i^{(s)}$ is the value of $\rho_i$ at iteration $s$ and $V_{\rho_i}$ is the variance chosen to ensure appropriate chain mixing. The value $\rho_i^{(s+1)}=\rho_i^*$ is accepted with probability $\alpha_{\rho_i}$, where 
\begin{equation*}
\alpha_{\rho_i}=\min\left\{1,\frac{\pi(\Theta^*|\bm{x})f_N(\rho_i^{(s)}|\rho_i^*,V_{\rho_i})\mathbbm{1}_{\rho_1^{(s+1)}<\cdots<\rho_i^*<\cdots<\rho_k^{(s)}}}{\pi(\tilde{\Theta}|\bm{x})f_N(\rho_i^{*}|\rho_i^{(s)},V_{\rho_i})\mathbbm{1}_{\rho_1^{(s+1)}<\cdots<\rho_i^{(s)}<\cdots<\rho_k^{(s)}}}\right\},
\end{equation*} 
where $\Theta^*=\{\bm{\rho}_{<i}^{(s+1)},\rho_i^*,\bm{\rho}_{>i}^{(s)},\Theta_C^{(s)},\Theta_M^{(s)}\}$, $\tilde{\Theta}=\{\bm{\rho}_{<i}^{(s+1)},\rho_i^{(s)},\bm{\rho}_{>i}^{(s)},\Theta_C^{(s)},\Theta_M^{(s)}\}$, $\bm{\rho}_{<i}^{(s+1)}=(\rho_j^{(s+1)})_{j<i}$, $\bm{\rho}_{>i}^{(s)}=(\rho_j^{(s)})_{j>i}$, $\Theta_C^{(s)}\subseteq\{\bm{w}^{(s)},\delta_1^{(s)},\delta_2^{(s)},v^{(s)}\}$ denotes the remaining copula parameters, which depend on the considered copula, at iteration $s$ and $\Theta_M^{(s)}$ denotes all the marginal parameters at iteration $s$.

\item Sampling $\bm{w}$.

The vector of copula weights is proposed from a Dirichlet distribution $\bm{w}^*\sim D(V_{w}\bm{w}^{(s)})$, where $V_w$ is chosen to be equal to 50. So $\bm{w}^{(s+1)}=\bm{w}^*$ with probability $\alpha_{\bm{w}}$ equal to
\begin{equation*}
\alpha_{\bm{w}}=\min\left\{1,\frac{\pi(\Theta^*|\bm{x})f_D(\bm{w}^{(s)}|\bm{w}^{*})}{\pi(\tilde{\Theta}|\bm{x})f_D(\bm{w}^{*}|\bm{w}^{(s)})}\right\},
\end{equation*}
where $\Theta^*=\{\bm{\rho}^{(s+1)},\bm{w}^*,\Theta_C^{(s)},\Theta_M^{(s)}\}$, $\tilde{\Theta}=\{\bm{\rho}^{(s+1)},\bm{w}^{(s)},\Theta_C^{(s)},\Theta_M^{(s)}\}$, and $\Theta_C^{(s)}\subseteq\{\delta_1^{(s)},\delta_2^{(s)},v^{(s)}\}$.

\item Sampling $\delta_1$ (skew-Normal and skew-T).

Since $\delta_1\in(-1,1)$, the proposal kernel is taken as the truncated Normal
\begin{equation}
\label{eq:mcmc1}
\delta_1^*|\delta_1^{(s)}\sim \mathcal{N}(\delta_1^{(s)},V_{\delta_1})\mathbbm{1}_{-1+\epsilon,1-\epsilon},
\end{equation}
where $V_{\delta_1}$ is the variance of the proposal distribution chosen to ensure chain mixing. So $\delta_1^{(s+1)}=\delta_1^{*}$ with probability $\alpha_{\delta_1}$ equal to
\begin{equation}
\label{eq:mcmc2}
\alpha_{\delta_1}=\min\left\{1,\frac{\pi(\Theta^*|\bm{x})f_{N(-1+\epsilon,1-\epsilon)}(\delta_1^{(s)}|\delta_1^*,V_{\delta_1})}{\pi(\tilde{\Theta}|\bm{x})f_{N(-1+\epsilon,1-\epsilon)}(\delta_1^{*}|\delta_1^{(s)},V_{\delta_1})}\right\},
\end{equation}
where $\Theta^{*}=\{\bm{\rho}^{(s+1)},\bm{w}^{(s+1)}\delta_1^*,\Theta_C^{(s)},\Theta_M^{(s)}\}$, $\tilde{\Theta}=\{\bm{\rho}^{(s+1)},\bm{w}^{(s+1)}\delta_1^{(s)},\Theta_C^{(s)},\Theta_M^{(s)}\}$, $\Theta_C^{(s)}=\{\delta_2^{s},v^{(s)}\}$ and $f_{N(-1,-1)}$ denotes the density of a Normal truncated in $(-1,1)$.

\item Sampling $\delta_2$ (skew-Normal and skew-T).

The proposal and acceptance of $\delta_2$ follows the same steps as in equations (\ref{eq:mcmc1})-(\ref{eq:mcmc2}) by substituting $\delta_1$ with $\delta_2$ and defining $\Theta^{*}=\{\bm{\rho}^{(s+1)},\bm{w}^{(s+1)}\delta_1^{(s+1)},\delta_2^*,\Theta_C^{(s)},\Theta_M^{(s)}\}$, $\tilde{\Theta}=\{\bm{\rho}^{(s+1)},\bm{w}^{(s+1)}\delta_1^{(s+1)},\delta_2^{(s)},\Theta_C^{(s)},\Theta_M^{(s)}\}$ and $\Theta_C^{(s)}\subseteq\{v^{(s)}\}$.

\item Sampling $v$ (T and skew-T).

For $v\in\mathbb{R}_{+}$, $v^*$ is proposed from a gamma $G(v^{(s)},(v^{(s)})^2/V_v)$, where $V_v$ is the variance of the proposal distribution chosen to ensure chain mixing.  So $v^{(s+1)}=v^*$ with probability $\alpha_v$ equal to
\begin{equation*}
\alpha_v=\min\left\{1,\frac{\pi(\Theta^*|\bm{x})f_G(v^{(s)}|v^{*},(v^{*})^2/V_v)}{\pi(\tilde{\Theta}|\bm{x})f_G(v^{*}|v^{(s)},(v^{(s)})^2/V_v)}\right\},
\end{equation*}
where $\Theta^*=\{\Theta_C^{(s+1)},v^*,\Theta_M^{(s)}\}$, $\tilde{\Theta}=\{\Theta_C^{(s+1)},v^{(s)},\Theta_M^{(s)}\}$ and $\Theta_{C}^{(s+1)}\subseteq\{\bm{\rho}^{(s+1)},\bm{w}^{(s+1)},\delta_1^{(s+1)},\delta_2^{(s+1)}\}$.

For $v\in\mathbb{N}$, $v^*$ is proposed from a discrete uniform distribution in $\{v^{(s)}-2,v^{(s)}-1,v^{(s)},v^{(s)}+1,v^{(s)}+2\}$. So $v^{(s+1)}=v^*$ with probability $\alpha_v$ equal to
\begin{equation*}
\alpha_v=\min\left\{1,\frac{\pi(\Theta^*|\bm{x})}{\pi(\tilde{\Theta}|\bm{x})}\right\}.
\end{equation*}

\end{itemize}

\section{BIC and DIC scores}
\label{appendixC}
\begin{table}[h!]
\centering
\scalebox{0.65}{
\begin{tabular}{l|cc|cc|cc|cc|cc|cc|cc|cc}
 & \multicolumn{2}{c|}{2G} & \multicolumn{2}{c|}{SN} & \multicolumn{2}{c|}{MO} & \multicolumn{2}{c|}{BL} & \multicolumn{2}{c|}{2T} & \multicolumn{2}{c|}{ST} & \multicolumn{2}{c|}{AL} & \multicolumn{2}{c}{CA} \\
 \hline
 & BIC & DIC & BIC & DIC & BIC & DIC & BIC & DIC & BIC & DIC & BIC & DIC & BIC & DIC & BIC & DIC \\
G1 & 9998 & 9860 & 9458 & 9424 & \textbf{9342} & 9255 & 9095 & 9008  & 10012 & 10175 & 10846 & 10705 & 10501 & 10472 & \textbf{8923} & 9072 \\
G2 & 9973 & \textbf{9604} & NA & NA & NA & NA & NA & NA & 9866 & 9997  & 10832 & 10333 & NA & NA & 8972 & \textbf{8928} \\
T1 & 9884 & 9657 & \textbf{9404} & 9489 & 9390 & \textbf{9190} & 9105  & \textbf{9004} & 9900 & 10007 & 10774 & 10434 & \textbf{10492} & \textbf{10387} & 8953 & 9078 \\
T2 & \textbf{9668} & 9635 & NA & NA & NA & NA & NA & NA & \textbf{9882} & \textbf{9983}  & NA & NA & NA & NA & NA & NA \\
SN1 & 10050 & 9693 & 9609 & \textbf{9389} & 9367 & 9324 & \textbf{8988} & 9024  & 10064 & 10226 & 10279 & \textbf{9865} & 10561 & 10427 & 8938 & 9091 \\
SN2 & 9986 & 9612 & NA & NA & NA & NA & NA & NA & 9912 & 9991 & 10282 & 10010 & NA & NA & 8988 & 8932 \\
ST1 & 9718 & 9632 & 9466 & 9446 & 9355 & 9260 & 9283 & 9157 & 9939 & 10065 & \textbf{10278} & 9999 & 10901 & 10402 & 8940 & 8934
\end{tabular}}
\caption{\footnotesize{BIC and DIC scores of mixtures whose components have non-zero wheights in our simulation study.}\label{table3}}
\end{table}

\begin{table}[h!]
\centering
\scalebox{0.65}{
\begin{tabular}{ll|cccccc}
 & &G1 & G2 & T1 & SN1 & SN2 &ST1\\
 \hline
 \multirow{2}{*}{River}& BIC & 39518 & 39497 & \textbf{39445}&39538 &39486 & 39518\\
 & DIC &39747 & 39618 & 39494 & 39896 & \textbf{39259} & 39593\\
 \hline
  \multirow{2}{*}{Leeds}& BIC & \textbf{7354} & NA & 7359 & 7367 & NA & 7370 \\
  &DIC& \textbf{7379} & NA & 7380 & 7382 & NA & 7384
\end{tabular}}
\caption{\footnotesize{BIC and DIC scores of mixtures whose components have non-zero wheights in our applications.}\label{BIC}}
\end{table}

\end{document}